# The Price of Empire:
# Unrest Location and Sovereign Risk in Tsarist Russia

Christopher A. Hartwell*
Professor of International Business Policy and Head of the International Management Institute
Zurich University of Applied Sciences (ZHAW-SML); Professor, Kozminski University
chartwell@kozminski.edu.pl

&

Paul M. Vaaler
Professor, John and Bruce Mooty Chair in Law & Business
University of Minnesota Law School & Carlson School of Management
vaal0001@umn.edu

**Abstract**

Research on politically motivated unrest and sovereign risk overlooks whether and how unrest location matters for sovereign risk in geographically extensive states. Intuitively, political violence in the capital or nearby would seem to directly threaten the state's ability to pay its debts. However, it is possible that the effect on a government could be more pronounced the farther away the violence is, connected to the longer-term costs of suppressing rebellion. We use Tsarist Russia to assess these differences in risk effects when unrest occurs in Russian homeland territories versus more remote imperial territories. Our analysis of unrest events across the Russian imperium from 1820-1914 suggests that unrest increases risk more in imperial territories. Echoing current events, we find that unrest in Ukraine increases risk most. The price of empire included higher costs in projecting force to repress unrest and retain the confidence of the foreign investors financing those costs.

THIS VERSION: November 13, 2023

*Corresponding author. This research has benefited from comments, criticisms, and suggestions from Franklin Allen, Edward Cook, Dmitry Didenko, Khurshid Djalilov, Scott Gehlbach, William Goetzmann, Tamar Gomez, Amanda Gregg, Sergei Guriev, Kristen Hartwell, Tarek Hussan, Guido Imbens, Michael Iselin, Andrew Karolyi, Aseem Kaul, Dmitrii Kofanov, Howard Lavine, Thorsten Lehnert, Timothy Lloyd, Arnold Lutz, Steven Nafziger, Timur Natkhov, Thomas Owen, Will Pyle, Koen Schoors, Myles Shaver, Andrei Shleifer, Pierre Siklos, Jason Smerdon, Sergei Smirnov, Laura Solanko, Rok Spruk, Karsten Staehr, Rolf Tschernig, Ishmael Tingbani, Daniel Treisman, Harald Uhlig, Gertjan Verdickt, Ilya Voskoboynikov, Joel Waldfogel, Alexander Wagner, Yongxiang Wang, Olha Zadorozhna, and Ekaterina Zhuravskaya. This research also benefited from presentations at the American Economic Association annual meeting, Bournemouth University, the European Economic Association annual meeting, the Swiss Society of Economics and Statistics annual conference, the Russia colloquium in Ghent, the WINIR conference in Utrecht, the University of Wisconsin, the University of Minnesota, and the Association of Slavic, Eastern European, and Eurasian Studies. We thank Kate Carlson at the University of Minnesota's U-Spatial group for help in visualizing our research with historic maps and state-of-the-art event mapping technologies. We thank Crédit Agricole's archivist, Pascal Pénot, and his colleagues, Nicolas Gueugneau Casa, Sibel Zora, and Cassandre Maubert, for giving us access to Crédit Lyonnais historical records of bank operations in the Russian Empire. We thank Mara Sybesma and Christopher Beach for research assistance. All errors remain our own.

## I. Introduction

*An empire founded by war has to maintain itself by war.* –Montesquieu (1734)

For at least three decades, academic researchers have connected increased sovereign risk with politically motivated, often violent unrest, threatening government debt repayment (e.g., Lee, 1993; Bekaert et al., 2014; Duyvesteyn et al., 2016). This literature has grounding in the real world: For example, in 2019, Bolivia saw sovereign credit rating downgrades and increased government borrowing costs after its presidential elections which led to a wave of riots, killing hundreds (Tuerk, 2019). Similarly, the Russian Federation's 2022 invasion of Ukraine has killed perhaps a hundred thousand or more people and caused the first default on Russian government bonds in more than a century (WEF, 2022). In the 2020s, the increasing frequency of riots, wars, and other forms of unrest continues to threaten sovereign creditworthiness and the cost of government borrowing across a range of developing and industrialized countries (Fitch, 2018; Allianz, 2021).

While the link between unrest and risk has been demonstrated, there has been almost no examination of how location matters as a mediating factor: is unrest which occurs closer to political power more threatening to the regime's ability to pay? It could be that the proximity of unrest to economic and political power centers interferes with productive activities (Kutan and Yaya, 2016) and threatens the actual procedures for making payments by disrupting the government (Castañeda and Vargas, 2012). On the other hand, an argument can (and will) be made that the need to project military power over vast distances is a greater issue for sovereign finances, stretching government budgets as they need to repress contentious territories. Research to date has neither explored this question nor investigated evidence related to either of these perspectives.

Part of the reason for this gap in the literature is that present-day country contexts for studying this research question are problematic. Poorer, less-developed countries may experience frequent and varied unrest events in different geographic locales but lack ready indicators of sovereign risk like sovereign credit ratings from major rating agencies or yields and credit spreads on government bonds trading in liquid financial markets. At the same time, wealthier and industrialized countries may have multiple sovereign risk indicators at hand but are less vulnerable to frequent and varied unrest events. No matter their level of development, present-day countries permitting some electoral competition and peaceful protest complicate the observation and analysis of unrest because they provide alternative legitimate means for individuals to voice political dissent (Campante and Chor, 2012). The would-be assassin or saboteur might instead dissent via the ballot box, the soap box, or Twitter/X.

However, one country context from the past avoids these research challenges and permits a "cleaner" study of unrest location and sovereign risk. Before its two revolutions occurring over the course of 1917, the Russian Empire was ruled by autocratic tsars brooking no legitimate political dissent. For most of this era, political parties and trade unions were outlawed and the state's censorship of news and opinion was comprehensive. As Figure 1 illustrates, this approach generated large amounts of political unrest in the capital city of St. Petersburg and other locations in and around the historic Muscovy Duchy, as well as thousands of kilometers away in distant, often recently conquered imperial territories in the Caucasus and Ural Mountains, on the Central European and Asian Steppes, and by the Black and Caspian Seas. There were attempted and successful assassinations of tsars and tsarist officials, peasant revolts and dissident rebellions in the countryside, riotous worker strikes and student demonstrations in cities, mutinies by military and naval units throughout the country, and armed conflicts against foreign countries in wars of territorial conquest or defense. Unrest was essentially the *only* means of dissent from government policies and brutal repression was the government's typical response (Pipes, 1974).

----------Insert Figure 1 about here----------

In contrast to these seemingly backward, autocratic politics prompting frequent and varied unrest, the same era saw Russia develop sophisticated international financial policies relying substantially on foreign investors buying and trading Russian government bonds and derivative securities in Amsterdam, Berlin, London, New York, Paris, and Vienna. Starting with Catherine the Great in 1769 (Ukhov, 2003), successive regimes sold bonds abroad primarily to fund Europe's largest standing army and to finance the acquisition, pacification, and defense of new taxable territories and subjects (Siegel, 2014). According to Ukhov (2003), by 1913, the value of Russian government debt traded abroad exceeded RUB 8 billion (equivalent to approximately USD 155 billion in 2022). Even with this staggering external debt, Russian government bonds were generally considered to be a "safe risk" for foreign individuals and institutions seeking higher returns and liquidity (Flandreau, 2013).

This paper exploits this historical country context to understand how unrest affects sovereign risk and how the geographic location of unrest changes those effects. Hand-collecting data on unrest events from primary sources such as Russian state bulletins and secondary sources such as contemporary compilations of political assassinations by foreign analysts, we create a database of unrest events occurring throughout the Russia Empire from 1820-1914. These unrest events are first categorized by type and then classified as either individual, collective, or external unrest event events; next, we distinguish the location, noting whether they occur in either the Russian homeland territories as they existed in 1820 or in the more remote imperial territories running from Warsaw in the west to Baikal in the east and from Siberia in the north to Chechnya in the south.

This data on unrest events is paired with data on two indicators of sovereign risk from the same era: Russian government bond yield percentages (yields) and percentage point spreads above yields on benchmark British government bonds (spreads). To isolate the effects of political unrest on sovereign debt, we use a two-pronged approach: first, we employ an event study methodology to capture the abnormal returns around events in our specific categories, including assassinations, assassination attempts, and collective unrest. The benefit of an event study in this context is that it allows us to give a causal interpretation to the correlation between changes in bond yields and political unrest over a short timeframe, where the causality is unlikely to run in the other direction (Gürkaynak and Wright, 2013). However, given that volatility in such events may also be influencing the path of the market's response (a point noted in Batchelor and Manzoni [2006]), we expand the analysis using asymmetric component GARCH-in-mean (ACGARCH-M) estimation, allowing us to also test explicitly the effects of other plausible covariates in driving bond market responses during the same event window. Across both of these approaches, our results show that: 1) single instances of individual, collective, and external unrest increase sovereign risk; 2) sovereign risk also increases when more unrest of the same type occurs during the prior 11 months; and 3) analyzed alone or cumulatively, individual and collective unrest events occurring in imperial territories increase sovereign risk more than when occurring in Russian homeland territories.

These results hold after various changes in sampling, model specification, and estimation approaches, including accounting for possible sample selection issues regarding where and when unrest is more likely to occur. The basic story also does not change if we alter the way we measure unrest location (from categorical to integral) or if we compare unrest effects in specific imperial territories. For these and other variations, we find a consistent and strong difference in the effect on sovereign risk related to where unrest occurs geographically. Finally, and perhaps most interestingly, we find that the negative effect on bond yields from unrest in the imperial territories is strongest when unrest occurred in the territories comprising modern-day Ukraine. In what may be an echo of current events, Russian government bond yields and spreads for unrest events in imperial Ukraine were not only higher than for unrest events in Russian homeland territories, but higher than in imperial territories generally. The price of empire in Tsarist Russia included demonstrably higher costs in projecting force farther to repress unrest and retain the confidence of the foreign investors financing those costs.

Our study advances academic research and related professional practice and public policy debates on the consequences of unrest in at least three ways. First, we advance fill the aforementioned gap in the literature in finance and economics (Chesney et al., 2011; Boutchkova et al., 2012; Bocola, 2016; Chaieb et al., 2020), political science (Jensen and Young, 2008), and public policy (Hogetoorn and Gerritse, 2021) by analyzing the locational effects of unrest on political risk and demonstrating how those locational effects matter for sovereign risk. Second, this paper repurposes theories of conflict emergence and resolution (Buhaug and Gates, 2002; Greig et al., 2018) to develop alternative perspectives about where unrest poses a greater threat to sovereign creditworthiness. Finally, we demonstrate the current relevance of historical research in finance: in particular, our empirical exploitation of Tsarist Russia's unique combination of repression and sophisticated finance yields new insights on historical unrest and sovereign risk with implications for modern day states, including today's Russian Federation.

## II. Perspectives on Unrest Location and Sovereign Risk

### *A. Unrest and Sovereign Risk*

As noted, extant research provides little guidance for understanding the effects of unrest location on sovereign risk. In part, this follows from a definitional issue, namely the conflation of unrest with political instability (Cumming et al., 2016). Alesina et al. (1996) are illustrative of that tendency, describing political instability as an unexpected government collapse that could follow from a vote of no confidence and defeat in a peaceful snap election or eviction at bayonet point after a violent military coup. Both could increase sovereign risk, but for different reasons. Political instability creates the potential for change of government, leading to shifts in policy preferences (Ashraf, 2022) and financial priorities (including potentially moving away from favoring timely payment to foreign sovereign bond investors before, say, alleviation of poverty back home). Sustained political instability means a revolving door of changing governments and policy preferences leaving foreign sovereign bond investors guessing about the willingness of any one government to meet the same financial obligations its predecessor did.

Unrest, defined here as a manifestation of political violence targeted at the state itself and carried out either individually or collectively, works differently. Unrest increases sovereign risk through at least three channels, decreasing the government's ability to meet financial obligations and increasing sovereign risk (Gaillard, 2014). In the first instance, unrest can destroy productive assets, harming physical plants, property, and equipment and vitiating a country's industrial base (Galvao, 2001).

Beyond the obvious physical destruction, unrest can also threaten or cause actual harm to investors (Castañeda and Vargas, 2012), deterring private investors or forcing existing investors to flee to quality (Rother et al., 2016). As Collier (1999) notes, these same two channels have public dimensions. Unrest can also destroy public infrastructure and deter public officials from the collection and redistribution of taxes, tariffs, and other public revenues (Isham et al., 1997). Logically, this can stifle the collection of government revenues financing government outlays, including those to foreign sovereign bond investors (Rother et al., 2016).

A third channel runs from unrest to the diversion of government revenues. This diversion need not follow from a change in government policy priorities as, for example, a looming election might see a shift in government outlays to fund more consumer goods and garner votes (Block, 2002). Here, diversion follows from unrest's increased demands on internal security and national defense (Knight et al. 1996), especially when unrest persists and is widespread. This trend is magnified when unrest specifically targets government officials with assassination, prompting the diversion of government savings (Edwards, 1996), especially when unrest targets the most senior government officials (Hartwell, 2022).

Thus, single unrest events can increase sovereign risk, and the effects can be heightened in a chain of such instances. Cumulative unrest can have the effect of institutional degradation (Knutsen and Nygård, 2015), wearing down the sovereign and their resources, negatively affecting the government's ability to meet financial obligations. Collier (1999) provides the basic logic in the context of longer-term civil conflicts. Think of a country-wide wave of politically motivated assassinations, repeated strikes shutting down essential services in major cities, or an unchecked guerrilla insurgency taking control of remote provinces. Above and beyond the physical harm and damage in each individual instance, repetition undermines the ruling regime's legitimacy, that is, the regime's unquestioned authority (Gilley, 2006). Braithwaite (2010) describes the effects of cumulative unrest as a "contagion" spreading from a focal point of sustained violence to neighboring regions; a message of state incapacity to quell violence in their former homes incites others near their current homes to doubt state capacity and challenge state authority. These processes raise the costs of public governance: rule of law; political stability; regulatory quality; government effectiveness; input from the citizenry; accountability of politicians failing to heed that input; and control of corruption (Kaufmann et al., 1999). Most important for our purposes, they also increase the state's "institutional challenge" to make credible commitments (Ramamurti, 2003), including commitments of the timely repayment of debts owed to foreign bond investors. Sovereign risk thus increases with unrest because of the government's increasing ineffectiveness, no matter who is in charge and no matter what policies they put first.

### *B. Unrest Location and Sovereign Risk*

Unrest differs from political instability on another dimension germane to our study: geography. Political instability inherently occurs where governments sit: Washington and Windhoek, Tokyo and Teheran, and Berlin and Buenos Aires. Unrest can occur anywhere. For example, the 2019 post-election riots which increased sovereign risk in Bolivia occurred in the capital city of La Paz as well as hundreds of kilometers away in regional cities and rural areas. In Colombia, the insurgency raging since 1964 has mostly been located in the countryside and has forced the Colombian government for decades to divert money towards the military (Kutan and Yaya, 2016).

The possibility of geographic variability gives rise to differing perspectives about when unrest poses greater threats to sovereign risk. Research across several fields guides our intuitive sense that individual unrest events as well as chains of unrest cumulatively increase sovereign risk. By contrast, we find no direct research guides for predicting locational differences in unrest effects. At best, we have two competing perspectives derived from research on the economic and political geography of conflict resolution (Buhaug and Gates, 2002; Buhaug and Røy, 2006; Campante et al, 2019), which we characterize as the "proximity" and the "projection" perspectives.

#### *1. The Proximity Perspective*

The proximity perspective suggests that unrest events located in capital and surrounding territories pose a greater threat because they more likely can imperil core political and economic activities central to the generation of government revenue for the timely repayment of sovereign debt. Consistent with this perspective, Campante et al. (2019) demonstrate that incumbent political elites in undemocratic regimes are constrained by the threat of insurrection, which is greater when potential dissidents are located closer to the country's political center. Their findings help explain, for example, why undemocratic regimes might remove the seat of government from larger cities to more remote areas. Their findings also help explain why the likelihood of armed conflict with dissidents increases when events indicating their defiance – a strike or a mutiny – occur closer to the capital.

The location of more recent, notable terrorist incidents may also exemplify this perspective. Terror groups are attracted to political and economic centers because of their potential for greater economic disruption and associated publicity. Bardwell and Iqbal (2021) estimate that the costliest terrorist event of this century was Al-Qaeda's 2001 attack on the World Trade Center in New York, which topped USD 40 billion. The next costliest terrorist event, a 2014 massacre of civilians in Nineveh, Iraq, had a negative economic impact of *only* USD 4.3 billion. Unrest in these centers may also prompt spillover effects detrimental to the economy and the government revenue it generates for timely bond payment. Wang and Young (2020) analyze a "fear" mechanism apparently causing local investor risk aversion in urban financial hubs like New York after individual terrorist incidents and repeated incidents over time. Similarly, terrorism nearer to manufacturing centers induces greater managerial pessimism and risk aversion in the form of decreased research and development expenditures (Cuculiza et al., 2017).

Proximity perspective logic suggests two trends relevant to our study. First, unrest nearer political centers, particularly in undemocratic regimes, diverts government revenues to costly conflicts. Second, unrest nearer economic centers causes greater economic losses and renders investors less willing to take the risks necessary to rebuild those centers. Both trends magnify threats to government commitments to sovereign bond investors.

2. *The Projection Perspective*

An alternative projection perspective suggests that unrest in more remote border regions poses a greater threat to sovereign creditworthiness due to the need to project force to stop the unrest. Boulding's (1962) classic on military conflict outlines a state's "loss of strength gradient" (LSG), when longer supply lines are required to impose its will further away from the country's seat of power. Related reasoning from de Mesquita (1981) highlights the cost of projecting power in distant, unfamiliar terrain and cultures. Distance and unfamiliarity impose supply challenges and threaten the morale of troops tasked with such projection (Hulme and Gartzke, 2021). In fact, inaccessibility or remoteness may actually make a region more likely to rebel or otherwise challenge the central government, due to the calculation made by rebels that the cost will be too high for government to bear (Tollefsen and Buhaug, 2015). Decreased power to project over distance undermines negotiations intended to extract tribute or other indicators of subservience from distant versus more proximate internal dissident groups or bordering external powers (Scott, 2009). LSG can also imperil foreign trade and investment in more remote, border regions. Greater logistical challenges in projecting power over distance along with less negotiating leverage to extract concessions from distant rivals implies more contentious periphery territories costing more to control (Buhaug, 2010; Markowitz and Fariss, 2013) and demanding greater diversion of funds otherwise dedicated to other purposes such as timely repayment of sovereign debt.

Consistent with this logic, Buhaug and Røy (2006) document that unrest in less-democratic Sub-Saharan African countries more often arises in border regions where dissident internal groups or bordering country governments can challenge sovereignty more easily. Similarly, Scott (2009) shows how unrest and refusal to be subjected to a sovereign power was more prevalent in more remote uplands of Southeast Asian states compared to territories in and around capital cities in the region. Buhaug and Gates (2002) also note that remote regions are more vulnerable to unrest when they hold valuable natural resources that provide substantial government revenues. Unrest threatening those resources and revenues may prompt a stronger (and more costly) repressive response, diverting government revenues from other uses including the timely repayment of sovereign debt.

The logic of the projection perspective suggests two trends relevant to our study. First, unrest events located in more remote, often border territories are more likely to draw a repressive response from government, particularly if those territories also include resources generating substantial government revenues. Second, that response could be more costly due to the challenges of power projection over substantial distance and cultural divides. Both trends magnify threats to government commitments to sovereign bond investors.

## III. Russia in Historical Context

Various characteristics of Tsarist Russian politics and finance in the 19th and early 20th centuries present a relevant context for the study of these perspectives of the effect of unrest location on sovereign risk. This context includes an unbroken line of autocrats brooking no legitimate political dissent, inclined toward brutal repression, and intent on both the modernization of economic and political centers in Russian homeland territories as well as the expansion of imperial borders and the control of the natural resources that expansion brings. These political moves engendered a broad array of unrest events located in Russian homeland and imperial territories, with some isolated but others occurring in waves, with potential cumulative effects. At the same time, Russia had early warning indicators of sovereign risk in the form of yields and spreads on Russian government bonds, which traded in efficient, liquid foreign financial markets supplied with timely information by contemporary standards.

### *A. Tsarist Russian Politics and Unrest*

It is difficult to over-emphasize the tsar's political authority in the 19th and early 20th centuries. As Pipes (1974) and others (e.g., Davidheiser, 1992) have noted, at a time when most European polities were transitioning to constitutional monarchies or republics with some popularly elected representation, Russia remained reliant on a totalitarian autocrat (*samoderzhets*) to serve as a bulwark against fears of political dominance by opportunistic foreign invaders and political chaos among bickering local nobles. A bureaucracy of courtiers seeking attention, influence, and landed titles carried out the Tsar's policies, with St. Petersburg constituting the political center of the Tsarist Russian

Empire throughout this period. Moscow and the surrounding region industrialized earlier and served as the regime's transportation and military hub (Cheremukhin et al., 2017).

Hosking (1973) contends that, at least from the mid-18th century, the tsarist system benefitted financially and politically from extending imperial boundaries. New exportable resources such as grain, iron ore, and gold paid an increasing share of court budget outlays, while new taxable land and the serf, peasant, and craftsman "souls" tied to it paid an increasing share of the costs for a standing army ranging from approximately 300,000 when first constituted by Tsar Peter II (the Great) in 1721 to more than 6 million in 1914. Land and resource grants from tsars placated an older class of landed nobles (*boyars*) and rewarded a newer, fast-growing class of courtiers (*dvoriane*) comprising the imperial bureaucracy. Territorial acquisitions in Central Europe during the late 18th century, the Caucasus and Far East during the 19th century, and the Balkans during the early 20th century shored up the regime financially and politically. But the same territorial acquisitions also created an increasingly expansive state to administer and defend against dispossessed, often hostile neighboring countries and, more troublingly, internal opponents (Pintner, 1984).

Autocratic regimes allow little political opposition and Tsarist Russia was no exception. For most of the 19th and early 20th centuries, political parties and unions were outlawed. State censorship was pervasive. Free speech and assembly were non-existent. As Bradley (2002) and Nafziger (2011) note, the few civil society organizations, publications, and programs of action emerging in the late 19th century served as conduits for pro-Tsarist Russian nationalism rather than political dialogue and dissent. Even the two great political reforms of this period, Tsar Alexander II's emancipation of serfs in 1861 and Tsar Nicholas II's convening (under coercion) of a popularly elected parliament (*Duma*) in 1906, did little to create legitimate channels for political dissent. Self-governing local councils (*zemstvo*) created after emancipation slowly collapsed due to lack of capacity or were co-opted by Alexander's bureaucrats to increase tax revenues. A little more than a year after its initial convening, Nicholas dissolved the parliament and altered electoral laws to ensure the election of more pro-government deputies.

With no legitimate means of political dissent, unrest became the principal means of indicating opposition to policies of the "father of the fatherland." Miller (2013: 59-60) notes that peasant rebellions in rural areas were "endemic" and caused "enormous devastation." Unwilling to countenance threats to their authority, successive tsars deployed the army and secret police (*Okhrana*) to be "ruthless in its repression." The growth of city-based factories in the late 19th and early 20th centuries prompted the emergence of an industrial working class and occasional worker strikes.

Infiltration of worker ranks by the secret police and then assault by army units became "the government's basic approach to the 'settlement' of industrial disputes." The same repressive combination quelled often riotous student demonstrations against the perceived inadequacy of regime reforms such as the abolition of serfdom in 1861. According to Bushnell (1985), the army itself was prone to unrest during leadership transitions and military setbacks. Mutinous armed forces prompted bloody suppression by loyalist units, followed by mass exiles or executions, whether it be in response to the Decembrist revolt of 3,000 soldiers against the newly crowned Tsar Nicholas I in the Senate Square of St. Petersburg in 1825 or in response to a series of mutinies against Tsar Nicholas II across the Empire in 1905 after defeats on land and sea during the Russo-Japanese War of 1904-1905.

Unrest in Tsarist Russia also took the form of individual acts of political assassination. The late 19th and early 20th centuries saw assassination waves perpetrated by outlawed political parties like the Socialist Revolutionaries, underground organizations like the People's Will, and "lone wolf" dissidents against tsarist officials, private bankers, and business executives closely associated with the regime. For Kotkin (2014: 61), "few options for political discourse existed, other than tossing a 'pomegranate' at an official's carriage and watching the body parts fly." According to Strakhovsky (1959: 357), "no fewer than 1,231 officials and 1,768 private persons were killed, and 1,284 officials and 1,734 private persons were wounded" in a wave of assassination attempts across Russian cities in 1907. Tsar Alexander II's successful assassination in 1881 by members of the People's Will may be the best-known example, but the aftermath of an unsuccessful attempt on the life of Tsar Alexander III in 1887 may have had greater significance in Russian history. Government retribution against the assassins led to the hanging of Aleksandr Ulyanov and the radicalization of his younger brother, the future Vladimir Ilyich Lenin.

Unrest as armed conflict had both internal and external dimensions in Tsarist Russia of the 19th and early 20th centuries. Many "internal" conflicts occurred on territories acquired from neighboring countries like Ottoman Turkey in the south, Sweden in the north, Austria in the west, and various khanates in the east. Khodarkovsky (1992) notes that leaders in newly conquered territories might agree to be "protected" by the tsar believing they were concluding a treaty between equals. But according to Suny (2001: 37), "Russians uniformly mistranslated the agreement as one of an inferior's supplication to the Russian sovereign." Such an approach fomented all manner of rebellion in imperial territories ranging from Poland to the Caucasus and Central Asia. Each instance drew a military response often aimed at putting down the rebellion and deterring others in an expanding empire from rising in rebellion.

## B. *Tsarist Russian Finances and Sovereign Bonds*

Tsarist Russia's political system may have seemed repressively backward compared to Western European states, but Tsarist Russia's approach to financing that system abroad was not. Ukhov (2003) and others (e.g., Flandreau and Zumer, 2004; Hartwell, 2021) explain how finance ministers serving the monarchy from Tsarina Catherine II to Tsar Nicholas II helped create a broad base of foreign investors ready to purchase and trade Russian government bonds and related derivatives. They were denominated first in Russian rubles and Dutch guilders but later in other currencies, while some bonds were fixed while others had flexible redemption periods. Additionally, they might be rolled over into new issues or exchangeable for precious metals such as gold to hedge against inflation and currency depreciation. Flandreau (2013: 20) concludes that these and other types of Russian sovereign bonds traded abroad were considered a "safe risk" among foreign investors notwithstanding at least two technical bond defaults in 1839 and 1885 as well as threats of foreign debt repudiation by left-wing dissidents seeking to topple the regime after 1905.

Ukhov (2003) explains how Tsarina Catherine II's finance ministry issued the first Russian sovereign bonds abroad in 1769. At the time, no Russian banks operated in prominent money centers like London, Amsterdam, or Paris. Many potential foreign investors lacked familiarity with Russia's economy, finances, and budgetary priorities. The finance ministry innovated around these obstacles by paying 6.5 percent of the bond issue's value to Hope & Company, an investment bank in Amsterdam. Hope & Company then purchased ruble-denominated sovereign bonds and issued guilder-based derivative securities under the bank's name collateralized by the Russian finance ministry's promise of regular ruble-based debt repayments. Thus, the Russian government borrowed its first 500,000 guilders (RUB 350,000) in 1769. By the time Catherine died in 1796, similar transactions in Amsterdam had increased Russian sovereign debt to 62 million guilders (RUB 41.5 million). From then to the 1840s, successive bond issues using the same two-step process occurred in Amsterdam through Hope & Company and later in London through investment banks like Barings and Rothschild.[1]

[1] Notes in an 1878 financial study (*étude financière*) produced for Crédit Lyonnais, a prominent French bank doing business in Russia, tell a slightly different story about the origins of Russian borrowing abroad. According to the Crédit Lyonnais study, from the 1760s, Catherine's government took out loans from unspecified lenders in Holland and the House of Wolf in Anvers, France. Her government also assumed part of the sovereign debt of the Kingdom of Poland when it ceased to exist after partition by Russia, Prussia, and Austria in the 1790s. Russia then consolidated these accumulated sovereign debts of more than RUB 46 million with its "first" loan of the same amount in 1798 from Hope & Company paying five percent and with a term of 17 years (Heinrich, 1878). Narratives by Ukhov (2003) and Crédit Lyonnais (Heinrich, 1878) concur about the importance of Hope & Company as the primary agent for Russian sovereign loans in the first half of the 19th century.

The move towards international finance was intimately connected to financing the regime's broader political aims (Ukhov, 2003). Until the 1840s, sovereign bond proceeds went exclusively to finance Russia's large standing army and its campaigns of territorial conquest, defense, and repression, with land-related tax revenues from boyars and courtiers helping to repay the bonds. In 1842, bond proceeds were used for the first time to finance infrastructure in the form of a railroad linking St. Petersburg with Moscow (Flandreau, 2013). By 1851, Tsar Nicholas I had established a separate railway troop service to protect transportation critical to military campaigns. By the 1890s, loan proceeds routinely went to fund state railroads, including the massive trans-Siberian railway project. As Pryadko (2020:2) explains, the resulting system made Moscow the Russian Empire's "major railway hub and a military center."

From the earliest foreign bond issuances, proceeds perpetuated a cycle: 1) government borrowing through sovereign bond issuance to fund an expanding military and later a related military transportation system; 2) territorial acquisition by that expanding military; 3) defense of new borders against dispossessed, sometimes hostile neighboring countries and the sometimes brutal repression of the dissidents inside those expanded borders; 4) allocation to and taxation of newly-acquired territories by boyars and courtiers; and 5) sovereign bond repayment by a government now overseeing a more geographically extensive state.

This foreign-financed military infrastructure protected Russia's expanding borders and permitted the exploitation of Russia's exportable resources. Conquered imperial territories in Ukraine and the Pontic-Caspian steppe comprised Russia's granary, where agricultural commodities comprised over half of Russia's total exports by the early 20th century (Dobb, 1948). Unrest in these more remote but vital agricultural territories posed a threat to export income for Tsarist Russia in the 19th and early 20th centuries, just as the invasion of Ukraine by the Russian Federation in 2022 reduced regional exports. A similar analysis of more remote unrest locations and greater sovereign risk might also apply to the conquered territories of the Caucasus and Caspian Sea Basin. From at least Tsar Peter II in the early 18th century to Tsar Alexander II in the mid-19th century, Russia periodically fought Ottoman Turkey, Persia, and local tribal groups for control of this region and its resources. By the latter half of the 19th century, oil and gas exploited and exported by private foreign companies had become important sources of government royalty revenue (van der Leeuw, 2000). Given its importance then (and now) for Russian finances, local unrest such as assassinations of local government officials or rebellions by local tribal groups sparked campaigns of repression often "prosecuted with incredible savagery" (King, 2008: 15).

Government sensitivities to remotely located unrest by internal dissidents and border rivals were often paired with regime commitments to project overwhelming force in response. We have already described Tsarist Russia's cycle of borrowing abroad for campaigns of territorial conquest, allocation for taxation, and then loan repayment to foreign bondholders (e.g., Ukhov, 2003). This cycle required an increasingly large standing army and navy. As Pintner (1984) notes, Russia's slower rate of civilian infrastructure development – roads, railways, canals, and ports – meant more local allocation of military and naval units across increasingly remote locales. However, their movement to trouble spots was complex and costly.

After naval defeat at Port Arthur during the opening moves of the Russo-Japanese War in February 1904, Tsar Nicholas II sent his Baltic Fleet on an 18,000 nautical mile trip to Russia's Pacific coast lasting seven months and burning more than 500,000 tons of coal. The Japanese then sank most of it at the Battle of Tsushima in May 1905. Additionally, Russia had 80,000 troops stationed in Port Arthur and the surrounding Liaodong Peninsula when hostilities with Japan commenced, and initial military setbacks there prompted Nicholas to pour in reinforcements from European Russia. Soldiers travelled more than 4,000 kilometers from Moscow to Irkutsk in Siberia via the Trans-Siberian railway and then marched 1,800 additional kilometers to the Manchurian war zone. By the summer of 1905, reinforcements taking this nearly two-month journey were arriving at a rate of 30,000 each month (Connaughton, 2003), illustrating a determination to project force no matter the complexity and cost.

A wave of assassinations predating the revolution and cresting over 1904-1907 targeted business leaders and bankers in the private sector and finance ministry officials in the public sector. During this same period, the peasants, who used to make punctual tax payments, instituted tax boycotts. As one contemporary observer noted, the government "lost its head" and avoided forced tax collections for months lest they provoke a general uprising (Mavor, 1925: 309). Thus, the cumulative effect of military setbacks and political assassinations prompted a breakdown in tax revenue collection at a time when the regime was also seeking to borrow abroad to rebuild military and naval capacity in the run-up to 1914 and the outbreak of the First World War (Strachan, 2003).

Somewhat surprisingly, financial information regarding these events was well covered in the domestic and foreign press, offering timely price discovery and trading liquidity, albeit at a lag. Pipes (1974) explains that, until the early 19th century, the regime treated news as a state secret requiring review by government ministers before dissemination to the outer world. No matter where information on unrest and other political events relevant to foreign bond investors arose in the Russian Empire, it tended to travel first to St. Petersburg and then abroad via state-censored newspapers and uncensored foreign diplomatic dispatches. Prior to the arrival of the telegraph in the 1850s, most news traveled by post roads established by Tsarina Catherine II or by shipping from St. Petersburg. Generally, news in European Russia took one-to-three weeks to reach St. Petersburg and then another two-to-three weeks to reach Amsterdam and London where, according to Neal (1987), weekly and later daily newspapers published spot (cash) market and futures prices on Russian government bonds and derivative securities.

But political news could sometimes move more quickly even from remote imperial territories. On December 1, 1825, Tsar Alexander I suddenly died of typhus in Taganrog, a city located on the Sea of Azov more than 1,500 kilometers from St. Petersburg. By chance, a former British ambassador to Russia happened to be sailing near the city. His report avoided the usual communication channels running through St. Petersburg. News of Alexander's death reached Warsaw in about a week, London in 12 days, and Paris in 17 days.

Figure 2 illustrates the impact of such news on Russian government bond prices. The x-axis marks dates before and after Tsar Alexander I's death while the y-axis measures daily bond price highs for 5-year Russian government bonds traded in Amsterdam and quoted in British pounds. The sudden and unexpected removal of an autocrat previously meeting all foreign financial obligations might very well undercut sovereign creditworthiness and prompt a decrease in bond prices abroad. Russian government bond price highs in December 1825 follow this conjecture. As news of Alexander's death reached Warsaw on December 8, then London on December 13, and then Paris on December 18, bond price highs dropped from 98.5 to 96.25 and then to 94.75 pounds on December 21$^{st}$ before rebounding to 98.5 pounds on Christmas Eve. Foreign markets responded to new information as quickly as it could be communicated from the Russia Empire.

----------Insert Figure 2 about here----------

By the end of the Crimean War of 1853-1856, new information was communicated in days rather than weeks and foreign investors (mainly the French) were kept updated on financial information incredibly quickly. Historical price information on bonds issued abroad and domestically was available on a daily basis in international media, with agents permanently located in St. Petersburg and Moscow (as well as floating agents traveling to other parts of the Russian Empire), often embedded with government administrators or military units (Borodkin and Perelman, 2011). Thus, foreign financial market participants received information from a steadily broadening network of sources arriving at a steadily faster pace. By 1900, Western media with access to telegraphs and telephones could quicken that pace to hours rather than days, with the nature of the product (financial instruments) meaning that transactions could also accelerate (unlike the commodities which Juhász and Steinwender [2018] examined).

We can also understand the efficiency of foreign financial markets based on the information *withheld* from participants and the broader public. Figure 3 illustrates the same daily price highs for the same 5-year Russian government bonds traded in Amsterdam before and after the birth of Tsarevitch Alexei on August 12, 1904. Inter-generational continuity of the tsarist system depended on a male heir, so the announcement of Alexei's birth should have boosted confidence in the longevity of that system and its finances. Indeed, bond price highs did show a slight increase from just less than 96 pounds on Alexei's birthday to 96.25 pounds on September 3, the day of his Christening. But traders lacked knowledge of Alexei's hemophilia, a condition that nearly killed him at birth and would threaten him throughout his short life. If it were not for the Romanov family's decision to keep his condition a secret, bond prices would more likely have fallen rather than risen.

----------Insert Figure 3 about here----------

A key issue in any financial market is liquidity. Figure 4 illustrates the apparently high liquidity of foreign financial markets for the same 5-year Russian government bonds. Campbell et al. (2018) measure market liquidity for a given bond in a given year from 0 (low) to 1 (high) based on how many months information on bond returns is *not* available, that is, posted returns are zero. Thus, market liquidity for bond *i* in year *t* can be derived from the following equation: Bond Liquidity = 1 – [(Number of months bond has zero returns posted)/12]. With years on the x-axis and this 0-1 measure of bond liquidity on the y-axis, we can plot market liquidity trends for Russian government bonds traded in Amsterdam from 1788-1914. With a few exceptions, peacetime scores range from 0.8-1.0, indicating high market liquidity. Notable exceptions include 1839 when bond liquidity dipped in response to a major currency reform and then a series of financial panics in Russian railroad, banking, and industrial stocks. Other dips in bond liquidity corresponded with military campaigns associated with unsuccessful anti-Napoleonic coalitions in the 1790s and early 1800s, naval and military defeats in the Russo-Japanese War of 1904-1905, and costly interventions with no territorial gains in the First and Second Balkan Wars of 1912-1913. By contrast, Russian government bonds maintained high liquidity in the face of Napoleon's invasion in 1812 and the wars of territorial conquest against Ottoman Turkey later in the 19th century.

----------Insert Figure 4 about here----------

Timely price discovery was yet another foreign financial market trend favorable to Russian government bond investors. With years on the x-axis, the left-hand side y-axis of Figure 5 measures monthly yield percentages for the same 5-year Russian government bonds traded in Amsterdam and London from 1788-1914 while the right-hand side y-axis measures monthly spreads for the same bonds compared to benchmark 10-year British government bonds. Flandreau (2013) conveys, and trends in Figure 5 generally confirm, that these bonds typically offered yields of five-to-seven percent and spreads approximately two percentage points (200-basis points) above benchmark bond rates. A few actual or threatened armed conflicts account for most spikes deviating from those trends: Napoleon's occupation of Moscow in September-October 1812; Russia's suppression of military insurrection by Polish troops in Warsaw in October 1831; the outbreak of revolution in France in March 1848 toppling the French monarchy and threatening others in Europe; the outbreak of the Crimean War in March 1853; the outbreak of the Russo-Turkish War in September 1877; and Austro-Hungary's intended annexation of Bosnia-Herzegovina prompting Russian threats of war in August 1908.

----------Insert Figure 5 about here----------

## IV. Empirical Methods

*A. Data and Sampling*

We exploit this historical context to understand the general effects of unrest on sovereign risk and to understand more specifically how unrest event location may change those effects consistent with either the proximity or projection perspectives. To do so, we collect data from several sources. Yields and spreads are standard measures of sovereign risk in empirical research whether the context be sovereign risk in the run up to presidential elections in the Russian Federation in the 1990s (Block and Vaaler, 2004; Vaaler et al., 2005) or in the wake of politically motivated terrorism around the world in the 2000s (Procasky and Ujah, 2016). Indeed, Ferguson (2006) suggests that bond yields and spreads drew more attention as indicators of sovereign risk during our period of study since investors then had few alternative indicators of a regime's economic and financial health (e.g., current information on government budget balances, trade balances, and foreign currency reserves). Following this literature, we collect monthly data on percentage bond yield returns for 5-year Russian government bonds trading in Amsterdam and London. We compare those returns to percentage returns on benchmark British government consols to derive monthly percentage point spreads. Data for both are available from 1820-1914 from Global Financial Data (2022).

Data on unrest events from 1820-1914 are hand-collected from English, Russian, and other foreign-language primary and secondary sources described by Hartwell (2022) and substantially expanded for this study. Primary sources include eyewitness accounts of unrest events in newspaper articles and state bulletins (e.g., Gazette, 1869-1917). Secondary sources include scholarly historical narratives about specific unrest events (e.g., Pipes, 1974) as well as unrest event compilations from recent years (e.g., Clodfelter, 2017), the Soviet era (e.g., Valk, 1961a, b), and tsarist times (e.g., Perris, 1905). We cross-check unrest event information, including information on event location, using different public and archival sources. For example, a 1905 *étude financière* produced by the St. Petersburg branch of the French bank Crédit Lyonnais confirms the dates and locations of strikes, riots, and assassinations in 1905 noted in secondary sources (Crédit Lyonnais, 1905). Our data collection strategy ensures broad unrest event coverage including many unrest events overlooked in past and present English-language sources. The resulting proprietary database assembled for this study represents a substantial advance over previous empirical studies limited to a single type of unrest event (e.g., peasant rebellions) occurring over several years (e.g., Finkel et al., 2015). but draws heavily on them to create a more holistic picture of the different types of political violence and unrest prevalent in Tsarist Russia.

To be included in our sample, we first require that the unrest event involve a prominent political individual or issue motivating the event.[2] We next require that there be fatalities and some repressive response from the regime. Fatalities need not include the intended political target and repressive government responses could involve police, the military, or other state personnel.[3] Finally, we require that the unrest event receive coverage in media sources, indicating that information about the unrest event was available to foreign financial market players. Typically, that last requirement is met when we observe contemporaneous coverage of an unrest event in foreign media.[4]

These unrest events are classified by type and location. For type, we classify three separate types of unrest events: individual, collective, and external. Individual events are assassinations, either successful or attempted but unsuccessful. For either, we can identify a specific political target – say, Tsar Alexander II, assassinated in 1881, or Grand Duke Nicholas, against whom an unsuccessful assassination attempt was made in 1908. Successful and attempted but unsuccessful assassinations are coded separately as 0-1 dummies, taking the value of one in the month they occur and zero otherwise.

---

[2] This sampling requirement excludes unrest events associated with ethnic or religious motivations, meaning Tsarist-era pogroms against Jews in the Russian Empire are thus not sampled. This sampling requirement also excludes unrest events targeting minor public officials (e.g., local tax collectors) or local political issues (e.g., local tax collection).

[3] This sampling requirement excludes several attempted assassination events where neither the targeted public official(s) nor bystanders were killed. It also excludes unrest events that drew no immediate repressive response from the regime, such as brief tax "strikes" in cities and oblasts (regions) where higher-level tsarist officials took no repressive countermeasures.

[4] For example, we include as an unrest event the assassination of the Governor-General of Finland, Nikolai Bobikov, in June 1904. It was reported on the same day of its occurrence by the *Press Democrat*, a Santa Rosa, California-based newspaper with no connection to Finland or Russia. Contemporaneous publication of news about an unrest even from the Russian Empire in a small foreign media market indicated information was widely available to foreign financial market players.

Collective events include targets and outcomes affecting internal groups rather than individuals: internal revolts, strikes, demonstrations, riots, mutinies, or rebellions including students, workers, peasants, dissident groups, military, or naval units. To be included in the sample, we again require that these collective unrest events result in fatalities, although, in many cases, the exact number of fatalities is not available. We have a single 0-1 dummy for each of these collective unrest events taking the value of one in months when they start and potentially continue, and zero otherwise. This means that dummies for collective unrest events may take the value of one over several months, such as, for example, the Revolution of 1905, where the dummy takes the value one across several months from its start in 1905 to the cessation of the political violence in 1906.

Finally, external unrest events include groups often located on the periphery of the Russian Empire, such as bordering country regimes and dissidents living abroad. Again, these unrest events can also prove difficult to sort for empirical study. Declared wars with Western European countries were fought both within the Russia Empire as with the Crimean War of 1853-1856 and outside the Russian Empire as with the Russo-Turkish War of 1877-1878. With this in mind, we more simply define external events resulting in fatalities from armed conflict with sovereign nations or groups primarily located outside the Russian Empire. Typically, these are wars of conquest such the Russo-Turkish War of 1877-1878. But they also include two armed conflicts that occurred on Russian territory: the Crimean War and the Russo-Japanese War of 1904-1905.[5] Each of these external unrest events are coded as dummies taking the value of one in months when hostilities commence until the fighting is ended and a peace treaty is concluded.

Individual and collective unrest events are also classified by geographic location. Here, our principal classification approach is categorical. Individual and collective unrest events occur either in the Russian homeland or imperial territories. As the depiction of the entire Russian Empire in Figure 1 illustrates, the Russian homeland includes Moscow and lands of the historic Muscovy Duchy as well as St. Petersburg. We also include in Russian homeland territories lands east of St. Petersburg and Moscow near the city of Perm, which saw early economic development in mining and forestry. Characteristics of the Russian homeland correspond well to the core political and economic territories which Campante et al. (2019) assume to be more likely to draw costly armed responses to unrest in undemocratic regimes.

[5] Both involved defense of previously conquered imperial territories against foreign invaders.

More distant imperial territories also include important urban centers such as Warsaw in Congress Poland or Kyiv in modern-day Ukraine. However, Russia's imperial territories more typically exhibited lower population, rail, and road density as well as slower industrialization rates in the 19th and early 20th centuries. On the other hand, imperial territories to the south and east include valuable natural resources: oil in the Caucasus; gold in Uzbekistan; wheat in Ukraine and the Pontic-Caspian steppe; and warm-water ports at Sevastopol on the Black Sea and Port Arthur on the Yellow Sea. These characteristics correspond well to the more distant and peripheral territories which Buhaug and Gates (2002) assume to be more likely to draw costly armed responses when challenged for control by internal groups and neighboring foreign rivals.[6] With this classification approach, we can compare, for example, changes in yields and spreads after the assassination of a government official in Moscow with the same changes after the assassination of a government official outside the Russian homeland in Kyiv, Warsaw, or Baku.

Historical sources have noted that literally thousands of assassinations occurred in the early 20th century alone (Ulam, 1977) while peasant rebellions also numbered in the hundreds (Finkel et al., 2015). This reality means that overlapping unrest events can and do occur in Russian homeland and imperial territories, dependent upon the category of unrest: for example, all individual unrest events in the form of attempted or successful assassinations as well as some cumulative events such as strikes begin and end in a single month, while other collective unrest events begin in one month and can persist over several months. Not surprisingly, perhaps the longest duration unrest event is external, as the Crimean War of 1853-1856 occurs over 28 months, a period when different forms of political violence occurred elsewhere within the Russian homeland and the imperium. Similarly, during the 20 months when the external unrest dummy for the Russo-Japanese War of 1904-05 takes the value of one, there are also individual unrest dummies taking the value of one for assassinations and attempted assassination events and collective unrest.

Our approach to coding these events means it may be more precise to define our dummies as "event-months," in that we measure both the fact that an event occurred and also how long it persisted. This approach yields a total of 551 individual, collective, and external unrest event-months which begin in or continue in one of the 1,125 months from 1820-1914; 208 unrest events are classified as individual or collective and occur in the Russian homeland, while 214 are classified as occurring in the Russian imperium and 129 are external unrest events occurring at or near the borders of the Russian Empire.[7] As Figure 1 again illustrates, unrest events are distributed across the Russian Empire with more in Russian homeland than in imperial territories. This distribution of unrest events between Russian homeland and imperial territories is not surprising given the higher population and economic densities as well as the better media coverage of events in the Russian homeland. That said, we have substantial comparable individual and collective unrest events in both Russian homeland and imperial territories. A peasant rebellion near Voronezh in the Russian homeland in October 1847 is followed by a peasant rebellion near in western Ukraine in February 1848. Assassination of the governor-general of St. Petersburg in January 1907 is followed by assassination of the former governor-general of Kutaisi Province in Georgia in July 1908. We have a plausible empirical basis for assessing both the general effects of unrest on yields and spreads and the effect differences related to unrest location.[8]

[6] Results reported in our study are robust to reasonable redefinitions of Russian homeland and imperial territories. For example, we obtain consistent results when limiting Russian homeland territories to the ancient Muscovy Duchy *not* including St. Petersburg and lands surrounding Perm. These unreported results are available from the authors.

### B. *Models, Measures, and Tests*

In order to isolate the impact of various forms of political violence on sovereign debt markets, we rely on a multifaceted identification strategy. In the first instance, we use a multiple event study analysis, studying the reaction of bond yields around the various events coded above, in order to generate a causal interpretation of market reactions to political unrest. In the vein of Bessembinder et al. (2009) and Alexiadou et al. (2022), we examine the return to bond yields, calculated as

$$R_t = \ln\left(\frac{Yield_t}{Yield_{t-1}}\right) \qquad (1)$$

Unlike the returns examined in most event studies (i.e., of prices), a negative return on yields can be interpreted as being desirable, signaling a change in the direction of lower risk, while a positive return on yields is undesirable, showing an increase in risk.

[7] As will be shown below, the sample encompassing the Russian imperium itself ranges from 165 event months to 246, depending on the specific definition used.

[8] A complete data appendix of sampled events, including event type, inception date (and duration), location, and related references is available from the authors and is posted on ICPSR.

As in standard event study methodology, we utilize these returns to calculate both abnormal returns and cumulative abnormal returns using two separate models. First, we use a raw returns model, which derives abnormal returns (AR) as:

$$AR_t = R_t - \overline{R_i} \qquad (2)$$

Which states that abnormal returns at time *t* are calculated as their deviation from the mean of the overall timeseries of the bond yields.

As a robustness test, we also utilize the constant mean model, which takes the form:

$$AR_t = \overline{R_i} + \varepsilon_i \qquad (3)$$

Which assumes that returns should equal the overall mean of the time-series plus an additional disturbance term assumed to have a mean of zero (MacKinlay, 1997). The constant mean model often performs as well as more complex ones (Brown and Warner, 1985) but suffers from the fact that outliers are given equal weight, meaning that there may be higher variance in the calculation of returns.

In both models, cumulative abnormal returns (CAR) are then defined as:

$$CAR_t = \sum_{j=t_0}^{t} AR_t \qquad (4)$$

To ascertain the significance of the event in question in generating abnormal returns, we rely on a commonly utilized parametric test for significance, the Adjusted Patell t-test of Kolari and Pynnönen (2010). However, heeding the advice of Ederington et al. (2015) regarding bias in bond market event studies (especially due to event-induced variance), we also use the non-parametric generalized rank t-test (GRANKT) approach of Kolari and Pynnönen (2011) to see the significance of the various violent events in our database.

This event study approach, carried out over the categories of political events noted in Table 1 and above, returns an average effect, so we supplement this monthly approach with selected cases using daily percentage bond returns. Daily data on Russian bond markets is not available in digitized form, and thus we have digitized selected time periods around unrest events using two daily Amsterdam broad sheets consulted by foreign financial market players in the 19th and early 20th centuries: the Amsterdam Securities Magazine (*Amsterdamsch Effectenblad*) and the New General Securities Sheet (*Nieuw Algemeen Effectenblad*). Issues of these broad sheets are available for research purposes at the Capital Amsterdam Foundation (Capital Amsterdam, 2022) and we have collected and digitized for the first-time daily price highs, lows, and closing prices for specific 5-year Russian government bonds floated in Amsterdam.

In addition to returns, our daily frequency event studies also account for two measures of market volatility: realized day-to-day volatility and Brownian intraday volatility. Realized day-to-day volatility is the log of daily yield returns squared. Following Parkinson (1980) and Alizadeh et al. (2002), we treat Brownian intraday volatility as range-based to adjust for potentially inflated bias during periods of high volatility when market microstructure noise may be more pronounced. Thus, intraday volatility is the square of intraday yield returns multiplied by a variance estimator of 0.361 (Parkinson, 1980).

These daily event studies utilize a historical mean model (HMM) for comparison in computing abnormal Russian government bond yield returns against market returns of a year of trading days prior to the event, and we close this historical window at 30 and 20 days prior to event occurrence.[9] The event study then uses 3-day (-1, 1), 7-day (-3, 3), and 11-day (-5, 5) windows to calculate event occurrence yield returns for calculation and comparison with historical yield returns. As with the average effect event studies, the proximity (projection) perspective of unrest location suggests that unrest events in Russian homeland (imperial) territories generate larger cumulative abnormal returns compared to unrest events located in imperial (Russian homeland) territories.

Finally, as a lot can happen in a month's time, especially in 19th century Russia, we also check these event study results via an asymmetric component GARCH in mean (ACGARCH-M) model. The ACGARCH-M model allows us to control explicitly for conditional mean effects of differing types of unrest on yields and spreads as well as short- and long-term yield and spread volatility trends, while also including plausible covariates which could be influencing bond yields over the month. Importantly, unlike an event study, this model can show the cumulative effects of political unrest in a short span of time (rather than over the entire span of the data). Our model is thus:

---

[9] Results reported in our study are robust to change in historic yield returns more than 30 days prior to event occurrence. These unreported results are available from the authors.

$$Y_t = \mu + \gamma Y_{t-1} + \beta Unrest_t + \rho M_t' + \delta \sigma_t^2 + \varepsilon_t \qquad (5)$$

Table 1 provides detailed information for all terms in (5). $Y_t$ is one of two dependent variables measuring sovereign risk (*Sovereign Risk*): yields or spreads in month *t*. The $\mu$ term is an intercept. $Y_{t-1}$ is the one-month lagged value of the same sovereign risk measures. This catch-all control accounts for unspecified historical trends in yields and spreads, including previous effects related to unrest intensity from overlapping events. *Unrest* is a 0-1 dummy taking the value of one in month *t* when an individual, collective, or external unrest event is occurring, and $M_t'$ is a vector of specific national and international economic and political controls that might also affect yields and spreads in month *t*. The $\sigma^2$ term accounts for the volatility of yields and spreads in month *t*, making (1) GARCH-in-mean. Finally, $\varepsilon$ is a month *t* error term.

----------Insert Table 1 approximately here----------

We estimate versions of (5) with different formulations of *Unrest,* just as in the event studies. Initially, it is one of four 0-1 dummies taking the value of one when a single instance of some type of unrest occurs anywhere in the Russian Empire (*Attempted Assassination Empire*, *Successful Assassination Empire*, *Collective Unrest Empire*) or on the periphery of it (*External Unrest*). Positive coefficient estimates indicate increased yields or spreads in the wake of single unrest events consistent with our intuition that unrest generally increases sovereign risk. As noted above, we also want to see the cumulative effect of unrest events occurring over a short span of time – in this case, over a year – and so we fashion a cumulative unrest indicator. This means that the *Unrest* term in (1) becomes a count value of 0-12 with zero for no event, one for the focal unrest event itself – say, the assassination of the governor-general of St. Petersburg in January 1907 – and the count of the same type of unrest event (*Successful Assassination Empire*) in the previous 11 months. Thus, we can assess the impact of an assassination given a recent context of more or fewer previous assassinations throughout the Russian Empire. Since we already control for recent background unrest intensity via the lagged dependent variable, we can think of this 0-12 cumulative measure as an indicator of recent unrest extensiveness. Again, our intuition is that a focal unrest event occurring in the wake of more cumulative unrest of the same type increases sovereign risk.

As with the event study, of more interest is where these unrest events occur. To that end, we also re-estimate (5) after partitioning unrest terms by location. Consistent with the proximity (projection) perspective of unrest location, we expect more (less) positive coefficient estimates for different types of single unrest events in Russian homeland territories (*Attempted Assassination Homeland*, *Successful Assassination Homeland*, *Collective Unrest Homeland*) compared to coefficient estimates for the same type of single unrest event in imperial territories (*Attempted Assassination Imperial*, *Successful Assassination Imperial*, *Collective Unrest Imperial*).

Of course, sovereign risk may vary for reasons other than unrest, and $M_t'$ in Equation 5 is a vector of national and international factors explicitly accounting for these other reasons. Given the paucity of reliable information for financial market participants generally during this period (Ferguson, 2006), our list of plausible covariates is short: 1) a 0-1 dummy equaling one in months *t* when a tsar "changes" (due to death of his predecessor) (*Tsar Transition*); 2) monthly returns on gold traded in London (*Gold*) as a proxy for global economic conditions; 3) month *t* returns on the ruble against the guilder (*Ruble-Guilder*) as a proxy for Russian economic conditions; and 4) a measure of drought severity in month *t* in the western grain-growing region of the Russian Empire taken from Cook et al. (2020) and coded from 0-3 and with higher numbers corresponding to higher drought severity (*Drought*). The inclusion of economic factors is to proxy for possible income shocks which can both affect the sovereign's ability to pay, while, as Xu (2022) notes, financial sector movements have the ability to impact trade for decades, and thus it is important to control for trade separately, which we do here via the exchange rate. The transition of Tsars, lower gold returns, lower ruble returns relative to guilders, and more severe drought conditions are assumed to increase sovereign risk measured as higher yields and spreads.

In (5), the volatility term, $\sigma^2$, is included in the level equation, making it GARCH-in-mean. Along with the error term ($\varepsilon$), we treat volatility as contemporaneous with *Unrest*. $\sigma^2$ is a placeholder that may be transformed into $\ln(\sigma^2)$ or $\sqrt{\sigma^2}$ to improve model fit based on commonly used information criteria such as Akaike (AIC) or Bayesian (BIC). However modeled, the volatility term can be partitioned into longer- and shorter-term components, with the long-term volatility component modeled as:

$$q_t = \omega + \alpha(q_t - \omega) + \gamma(\varepsilon_t^2 - \sigma_t^2) + \theta_1 Z'_{1t} \qquad (6)$$

In (6), $q_t$ represents month-to-month long-run volatility in yields or spreads. It reflects effects from an unrest event as well as effects stemming from controls of vector $M_t'$ in (5). It converges to the time-invariant volatility level ω at a speed of $\gamma$. The closer the estimated value of $\gamma$ is to unity, the slower that the modeled long-term volatility reverts to the mean. Thus, we can capture the persistence of volatility in response to unrest events. Equation (6) also includes vector $Z'$, a set of exogenous variables influencing volatility. Here, it is represented by volatility that might arise from "official" unrest events like a change in tsar or "unofficial" unrest events such as the assassination of an advisor to a tsar.

The short-term volatility component is thus:

$$\sigma_t^2 - q_t = \beta_0(\varepsilon_t^2 - q_t) + \beta_1(\varepsilon_t^2 - q_t)d_t + \beta_2(\sigma_t^2 - q_t) + \theta_2 Z'_{2t} \quad (7)$$

Where $\sigma_t^2 - q_t$ represents month-to-month transitory volatility in sovereign bond yields or spreads. The *d* term captures asymmetric effects in the form of a 0-1 a dummy variable equaling one when unrest prompts a shock. If $\beta_1 > 0$, there is a leverage effect present in the model. Unrest can increase short-term conditional variance more than "positive" news can decrease it due to, say, an improvement in national or international economic conditions. As with the mean equation, the $Z'$ vector in (7) captures the effects of unrest as well as the full set of controls (on the assumption that factors such as drought or exchange rates may influence short-term volatility but not long-term volatility). We use the Student's T distribution to model unrest event effects unless information criteria suggest replacement with the generalized error distribution (GED).

## V. Results

### *A. Descriptive Statistics*

Table 1 provides descriptive statistics for 5-year Russian government bond yields and spreads from 1820-1914. Mean yields are 4.81 percent over this period while mean spreads are 1.65 percentage points above benchmark British government bonds. These means comport with previous research (e.g., Flandreau, 2013) observing that, for example, Russian government bond spreads in the 19th and early 20th centuries were typically two percentage points above benchmark rates.

Individual, collective, and external unrest events could and did prompt a sudden and sometimes substantial change in foreign investor assessments of sovereign risk. In April 1877, on the eve of Russia's declaration of war on the Ottoman Empire, yields briefly shot up to 7.250 percent and spreads to 4.113 percentage points. The start of collective unrest across Europe in March 1848 saw the same two indicators of sovereign risk spike briefly to 6.671 percent and approximately 3.000 percentage points. Yields increased to 4.351 percent with spreads at 1.631 percentage points in late 1906 after a wave of assassinations across the Russian Empire and in the wake of the political turbulence surrounding the 1905 Revolution.

The bar charts in Figure 6 depict mean effects of different types of unrest on the same 5-year Russian government bond yields and spreads. Here, we find the lowest mean yields (4.732) and spreads (1.450) for collective unrest events such as worker strikes. The highest mean spreads are for attempted assassination (1.806) and successful assassination (1.775) events, perhaps highlighting investor concerns about dissidents targeting specific regime officials rather than the regime more generally. External unrest events such as wars of conquest have the highest mean yields (5.050), but not the highest mean spreads (1.747). This may follow from simultaneous increases in benchmark British government bonds tied to the cost of Britain's participation in Russian wars, such as opposing Russia in the Crimean War and supplying arms to Russia's Ottoman (1877-1878) and Japanese (1904-1905) war opponents.

----------Insert Figure 6 approximately here----------

The bar charts in Figure 7 break out mean 5-year Russian government bond yields and spreads by unrest event type in the Russian homeland versus imperial territories. Consistent with the proximity perspective, we find that mean yields (5.245) and spreads (2.208) for attempted assassination events in Russian homeland territories are higher than mean yields (4.561) and spreads (1.499) for the same unrest events in imperial territories. But consistent with the projection perspective, we find higher mean yields (4.898) and spreads (1.797) for successful assassinations in imperial territories compared to mean yields (4.810) and spreads (1.757) for the same unrest events in Russian homeland territories. Also consistent with the projection perspective, we find higher mean yields (4.880) and spreads (1.809) in the wake of collective unrest events in imperial territories compared to mean yields (4.646) and spreads (1.414) for the same unrest events in Russian homeland territories. Given the higher frequency of collective unrest, we take these descriptive trends as preliminary evidence giving more support to the projection perspective on unrest location and effect differences.

----------Insert Figure 7 approximately here----------

### B. *Event Study Results*

The event study results described above are shown in Tables 2 (for the raw returns model) and 3 (for the constant mean model), ascertaining the effect on the return on bond yields for events happening anywhere in the Empire, and then broken out by their location. The first and most obvious result is that both models return similar effects across categories, with nearly identical coefficients and significance levels, apart from the Empire-wide event studies, which have little significance in the raw returns model. Indeed, the only variations can be found in significance levels between the parametric and non-parametric tests and, in the aforementioned case of the Empire-wide studies, between the raw return and constant mean model; given the possible issues with event-induced variance, our preferred approach would be to rely on the GRANKT test, but the adjusted Patell approach may also provide information on these effects. Turning to the specific impact of political unrest, across both models we can see that it has a differential impact on bond yields according to modality and, more importantly, lends credibility to the projection perspective. In each instance of unrest occurring in the imperium, bond yield returns increase in a statistically significant manner, while the results for unrest occurring in the Russian homeland are far more mixed: for example, attempted assassinations in Russia appear to decrease bond yield returns by 1.3% while attempted assassinations in the imperium increase bond yield returns by approximately 2%. This difference is most striking in terms of collective unrest, where collective violence in the Russian homeland has no significant effect on bond yields in every model but the constant mean one, using the GRANKT test (at the 10% level) but various permutations of collective unrest in the imperium all lead to significant and substantial increases in bond yields ranging from 0.1% to 1.1%. This implies an increase of approximately 0.05 percentage points in the level of bond yields for collective unrest in the imperium if we use the mean of yields across the whole sample (i.e., from 4.81% to 4.86%).

----------Insert Tables 2 and 3 approximately here----------

In addition to these results using monthly data, Tables 4 and 5 also report results from two event studies analyzing abnormal Russian government bond yield returns surrounding specific unrest events in Russian homeland and imperial territories, confirming the broader trends observed in Tables 2 and 3. Table 4 reports results from event study analyses of the CARs associated with the Obukhov Defense events of May 1901 in St. Petersburg, a workers' strike which morphed into a series of deadly clashes by workers and neighborhood residents with police and army units, leading to several fatalities and destruction of property. Marxist historians tout the Obukhov Defense as a seminal moment in the formation of the Russian workers' movement (Potolov, 1997). Given the strategic political importance of St. Petersburg and the emergence of a robust workers' movement there, we might expect foreign sovereign bond investor concerns about unrest and related sovereign risk to increase. However, the results in Table 4 undercut that expectation. CARs and related volatility in abnormal bond returns are assessed in and around May 1 when the strike began and May 9 when clashes with police and army units began, and we find no significant changes indicative of increased sovereign risk among foreign bond investors. By contrast, CARs in the narrowest three-day window in Column 1 suggest that sovereign risk significantly decreased rather than increased.

----------Insert Table 4 approximately here----------

This outcome can be contrasted with another series of clashes with police and army units 3,000 kilometers away in the Ukrainian port city of Odesa in February 1878. Then as now, Odesa was an important grain shipping center, crucial for the tsar in linking the Russian Empire to the West and providing the Russian Empire with important grain export income. Odesa was also a gathering place for political dissidents, including Ivan Kovalsky, a member of the People's Will and a dedicated revolutionary with a printing press distributing information critical of the tsarist regime. On February 11, 1878, police raided Kovalsky's apartment to arrest him and seize his printing press. According to Ulam (1977), Kovalsky first tried to shoot the arresting police, but his gun jammed, so he stabbed several of them. Kovalsky's followers started their own gun battle with the police, and the army was called in to quell the unrest.

----------Insert Table 5 approximately here----------

Given the strategic economic importance of Odesa, the exposure of a revolutionary cell leading to street violence there might well raise concerns among foreign sovereign bond investors. Results in Table 5 support this conjecture. We observe significantly increasing CARs and intraday abnormal bond yield return volatility across different event windows. The prospect of more costly unrest was real. Revolutionaries inspired by Kovalsky started an assassination wave during the remainder of 1878 ultimately killing the Mayor of Kyiv and the Chief of the Secret Police in Moscow. Foreign sovereign bond investors may well have anticipated the higher costs of repression associated with unrest events starting in Odesa and spreading across the Russian Empire.

### *C. ACGARCH-M Results*

The event study results provide evidence for the projection perspective but, as noted above, it is possible they are also capturing additional effects occurring at the same time as the political unrest, and thus a more flexible identification strategy is called for. Table 6 reports results from ACGARCH-M analyses of general unrest effects on 5-year Russian government bond yields and spreads. Columns 1-2 of Table 6 present results with only the controls as noted in (1). Coefficients for the lagged dependent variable and four controls exhibit the expected signs in both columns. In Column 1 analyzing yields, three of the four controls are statistically significant at commonly accepted levels of at least 10 percent ($p < 0.10$). There, for example, yields increase approximately 5.7 basis points in months when a tsar dies, as with Tsar Alexander I in December 1825. In Column 2 analyzing variation in spreads over benchmark British government bonds, two of the four controls are significant, and a one standard deviation appreciation of the ruble against the guilder decreases spreads by approximately 1.9 basis points.

Columns 3 through 10 then add our different unrest dummy variables to this baseline, thus permitting assessment of Russian Empire-wide effects of single unrest events on sovereign risk. Consistent with intuition that unrest increases sovereign risk, we find in seven of eight instances that unrest coefficients exhibit positive signs statistically significant at commonly accepted levels. In Column 4, we find that attempted assassination events increase 5-year Russian government bond spreads by approximately 5.0 basis points ($p < 0.01$).[10] In Column 10, we find that external unrest events increase spreads by approximately 2.6 basis points ($p < 0.01$). Both increases are economically substantial. The value of outstanding Russian government bonds in 1914 was USD 155 billion in 2022 dollars. Thus, a 2.6 basis point increase in debt costs translates into an additional USD 403 million in annual interest in 2022 while a 5.0 basis point increase in debt translates into an additional USD 775 million in annual interest in 2022. Overall, results in Columns 3-4 and 6-10 indicate substantial support for our intuition that single instances of individual, collective, and external unrest lead to statistically significantly and economically substantial increases in sovereign risk.

----------Insert Table 6 approximately here----------

ACGARCH-M results reported in Table 7 convey similar points regarding cumulative effects of unrest on sovereign risk. Across all eight columns, we find coefficients on cumulative unrest dummy variables that are positive at commonly accepted statistical levels. This is consistent with intuition that chains of similar unrest events prompt cumulative effects on sovereign risk in the forms of higher 5-year Russian government bond yields and spreads (although, perhaps surprisingly, there is little effect on yield and spread volatility). Column 3's coefficient for the cumulative impact of successful assassinations (*Cumulative Successful Assassination Empire*) on yields (0.012, $p < 0.01$) is both statistically significant and economically substantial. Based on these results, we infer that a year-long wave of assassinations like that across the Russian Empire in the mid-1900s could have increased yields by more than 14 basis points. This increase in borrowing costs would have come at a time when Russia was seeking foreign capital to rebuild armed forces destroyed in the Russo-Japanese War of 1904-1905 and accelerate industrialization in St. Petersburg, Moscow, and the Perm region (Siegel, 2014).

[10] Interestingly, successful assassinations appear to have the opposite effect, lowering spreads. This may be a function of the person targeted (i.e., if markets were not sanguine about a particular politician) or could be (more realistically) an uncertainty effect: that is, attempted assassinations may signal more volatility ahead, while a successful assassination – especially if the assailant is apprehended – can be seen as a one-off event.

----------Insert Table 7 approximately here----------

While the results in Tables 6 and 7 provide evidentiary support for our intuition about the general effects of unrest on sovereign risk, they tell us nothing about differences in effects related to unrest location consistent with either the proximity or projection perspectives. The results in Table 8 do this by partitioning unrest events by location either in Russian homeland territories such as in and around Smolensk (e.g., *Successful Assassination Homeland*) or in imperial territories such as in and around Baku (e.g., *Collective Unrest Imperial*).[11] Columns 1-6 of Table 8 simultaneously estimate effects on 5-year Russian government bond yields and spreads when a single unrest event occurs, while columns 7-12 do the same for cumulative unrest effects.

For single unrest events, the results are largely consistent with the projection perspective. Unrest events in more remote imperial territories prompt higher yields and spreads, suggesting that foreign bond provide information to investors about sovereign creditworthiness; indeed, unrest events in the imperial territories also lead to statistically significant elevated volatility in the short-term for the inception of single acts of collective unrest and especially for cumulative assassinations and collective unrest. Excepting the results from Column 4, we find that coefficients for all dummies representing unrest event types in imperial territories exhibit positive signs significant at either the five or one percent levels. Importantly, these estimates indicative of higher sovereign risk are both significantly different from zero and from the simultaneously estimated Russian homeland unrest event dummies.

----------Insert Table 8 approximately here----------

Columns 1-2 results for attempted assassination events in Russian homeland and imperial territories are emblematic of this evidence. The coefficient for the *Attempted Assassination Imperial* 0-1 dummy is positive and significant at the one percent level for both Russian government bond yields (0.059, $p < 0.01$) and spreads (0.082, $p < 0.01$). Coefficients for *Attempted Assassination Homeland* are positive for yields (0.039) and spreads (0.026), but their magnitudes are substantially lower and statistically insignificant. Moreover, coefficients for *Attempted Assassination Imperial* are significantly higher than their *Attempted Assassination Homeland* counterparts at the five percent level. Column 2's coefficient suggests that a single attempted assassination in imperial territories sometimes hundreds or thousands of kilometers from St. Petersburg or Moscow increases spreads by 8.2 basis points translating into millions of more rubles in 1914 and hundreds of millions of more dollars in annual interest in 2022.

[11] For these regressions, we split the difference with the collective unrest variables from the event studies and use the unrest variable including the Caucasus unrest but not including the Caucasus wars, which were a military campaign of conquest rather than political instability.

Columns 7-12 results exhibit the same pattern in the case of cumulative unrest locale and sovereign risk. Coefficients for our dummies representing three different types of unrest in imperial territories exhibit positive estimates significantly different from zero and from their Russian homeland territory counterparts. They again follow projection perspective assumptions that cumulative unrest in imperial territories raises sovereign risk more than cumulative unrest in Russian homeland territories. Higher costs of projecting power to suppress unrest in more contestable locales gives foreign bond investors greater concern about the Russian government's ability to meet its financial obligations.[12]

### *D. Robustness Tests*

1. Contagion and Spillovers

Within the event study and ACGARCH-M frameworks, it appears that the projection perspective dominates. However, there is a possibility that homeland and imperial events can be distinctly separated due to the contagious nature of unrest. While there was no organized common revolutionary movement across Russia during the 19th century, planning simultaneous attacks or unrest across the entire empire (similar to the reach al-Qaeda had in executing the 9/11 terror attacks), there was always a chance for opportunistic violence. For example, unrest in Ukraine could be closely linked to the actions of activists in other parts of Russia, making it difficult to events to a specific geographic location. In one sense, we have addressed this issue in the first set of ACGARCH-M regressions, using dummies if there was political unrest anywhere in the Russian Empire. But to further dig into this plausible ability, we augment the baseline regression of Equation 5 in two separate ways: one, by including a dummy if there were multiple events for that month across any events and two, by including what is essentially an interaction dummy, equaling one if there was any other event *and* the event which is the subject of that regression. For example, if the model examined the effect of successful assassinations, in addition to a dummy for successful assassinations in Russia and one for successful assassinations in the imperium, we would include a dummy equal to 1 if there was a successful assassination in Russia and any other type of unrest; similarly, this same model would also include a dummy equal to 1 if there was a successful assassination in the imperium and any other type of unrest. In this way, we can see the effects of contagion on bond yields.

[12] These ACGARCH-M model results prove robust to several variations in sampling, model specification, and estimation, with additional tests shown in the Internet Appendix.

----------Tables 9 and 10 here----------

The results of both of these approaches are shown in Tables 9 and 10 and they do not differ substantially in terms of results, either from each other or from the event studies or the baseline ACGARCH-M model. In particular, attempted assassinations in the imperial territories drive both yields and spreads higher significantly, while attempts in the Russian homeland have no impact; similarly, collective unrest in the homeland actually drives down both spreads and yields while in the imperium it results in much higher risk premia. The impact of multiple events is uniformly positive in Table 9, showing more than one event occurring simultaneously drives up both yields and spreads, while separating these effects out by location in Table 10 shows that (for the most part), it is the aggregation of imperial disturbances and other events which generate much higher borrowing costs for the sovereign. Finally, and interestingly, successful assassinations in the imperium appear to have a *negative* and significant effect on bond yields and spreads (Table 10, columns 3 and 4), contrary to Table 8, but when combined with another event of political unrest, they have a profoundly positive effect (equivalent to a 2% rise in bond yields from their mean over this timeframe). Thus, while the presence of other bouts of unrest may amplify uncertainty, apart from successful assassinations in the imperium, the effect of events farther away from St. Petersburg appears to be more pronounced in terms of increasing borrowing costs than events occurring in the Russian homeland. This again is in line with the projection perspective.

2. Measuring "Projection"

The event study and ACGARCH-M models defined unrest event location categorically either in Russian homeland or imperial territories, capturing distinctive historical and enduring cultural differences between conquering powers and conquered lands. This measure of distance also correlates well with integral distance from the political center of St. Petersburg to most unrest events. However, such effects could be better captured by an integral approach that explicitly measures distance and the costs of power projection that may increase with it. This would be difficult to model in an ACGARCH-M framework as, in months with no unrest, the integral measure of distance would drop out.

To deal with this challenge, we also employ a two-stage Heckman model with a different unrest locational measure, specifically the distance from St. Petersburg (the seat of power) in kilometers for each collective unrest event. For the first stage probit selection model, we identify several unrest event likelihood determinants from political science research (Blanco and Grier, 2009; Bowlsby et al., 2020; Buchheim and Ulbricht, 2020) to fashion a theory for the likelihood of an unrest event starting. The resulting first-stage probit model of unrest event likelihood is:

$$X_t = \mu + \gamma X_{t-1} + \rho M_t' + \delta A_t' + \varepsilon_t \quad (8)$$

In (8), $X_t$ is a 0-1 dummy taking the value of one when an unrest event occurs in month *t*. On the right-hand side of (4), we include an intercept (μ), a lagged dependent variable ($X_{t-1}$), the same vector of controls *M'* from (1), and an error term ($\varepsilon_t$). This specification captures a point which was of less importance in the earlier models but could be crucial when considering the effect of linear (or, rather, post road) distance: as Berman and Couttenier (2015) note, it is possible that economic shocks can affect the likelihood of conflict as well, and thus economic variables from (1) may capture this effect.[13] The right-hand side of (8) also includes a new vector of controls, *A',* that could help explain variation in the likelihood of an unrest event in month *t*. This new vector includes several terms derived from the literature: a 0-1 dummy taking the value of one in months *t* when there is serfdom (*Serfdom*), which should decrease unrest likelihood by decreasing the movement of peasant farmers tied to land; the percentage of annual court budget expenditures in months *t* going to the Ministry of Interior charged with state security and censorship[14] (*Interior*), which should decrease unrest likelihood given more resources to suppress it; and the log of rubles from annual exports of cereals in months *t* (*Cereals*), which should decrease unrest likelihood given more wealth to accommodate would-be dissidents.

The second-stage linear model (5) estimates variation in sovereign risk ($Y_t$) as a function of the right-hand side terms from (4) and *Distance*:

$$Y_t = \mu + \beta_1 Distance_t + \rho M_t' + \delta A_t' + \varepsilon_t \quad (9)$$

*Distance,* as noted above, is measured as the number of kilometers from St. Petersburg to the unrest event location using either actual historical post roads from the early 19th century (Atlas, 2022) or contemporary roads or rail lines often added to these historical post roads as the Russian Empire expanded in the later 19th and early 20th centuries. Given that the volatility of both yields and spreads were important in the AC-GARCH conditional mean equations, we also use specifications both with and without realized volatility as part of the Heckman process.[15] To capture effects resulting from temporal trends, we also include in one regression for each risk metric the number of years that a territory outside of the Russian homeland experiencing unrest was under Russian rule, on the supposition that territories which were subjugated longer would be easier to bring under control and thus have less effect on sovereign debt.[16] This measure is also interacted with the distance variable to see the specific effects of longer periods of subjugation in mediating the effect of distance.

---

[13] We believe that the effect of previous economic shocks on unrest is captured in Equation 1 by the lagged dependent variable.

[14] Spending on the Ministry of Interior increases almost every year from 1820-1914 as does the total court budget. At the same time, both exhibit different annual rates of increase apparently following trends similar to business cycles. With this in mind, we separate the cyclical component using a Baxter-King (1999) bandpass filter. We then use the transformed annual time series.

----------Table 11 here----------

The results of the 2-stage Heckman maximum likelihood (ML) approach are shown in Table 11. Both models including volatility and those without support the projection perspective, as the coefficient on *Distance* in (5) positively and significantly affects both bond yields (0.0001, $p < 0.01$) and spreads (0.0001, $p < 0.05$) in each instance. This result holds when controlling for bond volatility and is strengthened when the interaction term for years under Russian rule is included (Columns 3 and 6 of Table 11), with an increase of 0.0002 points ($p<0.001$) for each kilometer of distance. Put into perspective, each thousand kilometers of distance to an unrest event would increase bond yields and/or bond spread by 0.2 points, an increase of approximately 5% of the mean bond yield over this time period.

3. Unrest in Imperial Ukraine

In 2023, it is impossible to avoid comparison of historical unrest and sovereign risk trends in our study with similar trends related to the current war forced on modern day Ukraine by the Russian Federation. Not just political leaders, but many average citizens in Russian today assume that Ukraine lacks separate nationhood. For them, Russia's "special military operation" in Ukraine is not so much an invasion of another sovereign state, but reclamation of a wayward province with errant nationalist aspirations (Volkov and Kolesnikov, 2022).

With this current context in mind, it is interesting to consider whether similar attitudes informed similarly repressive Tsarist Russian policies toward Ukraine as an imperial territory in the 19th and early 20th centuries. Tsar Nicholas II's words certainly suggest a close fit with current Russian skepticism about Ukrainian nationhood: "There is no Ukrainian language, just illiterate peasants speaking Little Russian" (Peterson, 2014). Current Ukrainian views of a national identity separate from

[15] The AC-GARCH-M regressions model the volatility, as noted above, from the data. Here, we use a much simpler approach and include realized volatility (RV), where RV is calculated as the square of returns on either bond yields or risk spreads.

[16] Thanks to a commenter from the Swiss Economic Society annual conference who suggested this control variable.

Russia have deep historical roots at least from the perspective of Ukrainian writer Andriy Kurkov: "For Ukrainians who have never had their own tsar (we do not count princes and other local feudal lords), the Motherland, their homeland has always been more important than a foreign tsar and – which is the worse for Russia – more important than faith" (Kurkov, 2020: 86).

----------Insert Table 12 here----------

That historical tension about Ukrainian national identity might have also mattered to Russian government bond investors in foreign financial markets of the 19th and early 20th centuries. Table 12 reports results from an additional ACGARCH-M analysis of 85 collective unrest events in Russian homeland territories (*Collective Unrest Homeland*) versus 43 collective unrest events in the imperial territories of modern-day Ukraine (*Collective Unrest Ukraine*). Column 1 reports the effects of these unrest events on 5-year Russian government bond yields while Column 2 reports the same for spreads. The contrasting unrest effects are clear. Collective unrest in Russian homeland territories decreases rather than increases yields (-0.029, $p < 0.01$) and spreads (-0.018, $p < 0.01$), but collective unrest in Ukraine increases both measures significantly and substantially. Yields increase by approximately 6.7 basis points (0.067, $p < 0.01$) while spreads increase by 10.8 basis points (0.108, $p < 0.01$) with each unrest event.

Columns 3 and 4 of Table 12 compare sovereign risk effects for the same collective unrest events in the imperial territories of modern-day Ukraine to 33 other unrest events occurring in imperial territories outside Ukraine (*Collective Unrest Other Imperial*). The magnitude of the sovereign risk increases for collective unrest events in Ukraine is again larger. Results show that 5-year Russian sovereign bond yields increase by 5.1 basis points ($p < 0.01$) in Ukraine versus only 2.0 basis points ($p < 0.10$) elsewhere in imperial territories. Additionally, spreads connected with unrest events in Ukraine increase by 6.9 basis points ($p < 0.01$) while unrest events in other imperial territories show no significant effect on spreads.

----------Insert Table 13 here----------

Finally, to bring us full circle, we also utilize the event study methodology here to ascertain the effects of unrest in Ukraine on bond yield returns, shown in Table 13 and Figure 8. Using two windows – one month before and after and a longer window of four months beforehand to four months after - the results in Table 13 also point to a marked increase in bond yields. Under the tighter one-month window, the results are significant only for parametric testing, but over a longer term, bond yields are highly statistically significant and show an increase of approximately 3% in bond yields for a collective unrest event in modern-day Ukraine. It appears that foreign bond investors anticipated higher costs to the Tsarist Russian government from defiance by Ukrainians in the 19th and early 20th centuries. This is eerily similar to current events, as the Russian Federation is paying so much in both blood and treasure for having underestimated the resilience of the sovereign nation of Ukraine.

## VI. Conclusion

This paper has examined the question if unrest location matters for sovereign risk, developing alternative perspectives for understanding how geographic location could change the impact of unrest on a government's creditworthiness: a proximity perspective, assuming that unrest closer to political and economic centers in a country increases sovereign risk more, and a projection perspective, assuming that unrest in more remote areas nearer contestable borders or valuable natural resources in a country is more important. To assess support for either of these perspectives, we turned to the unique historical context of Tsarist Russia, which had frequent and varied forms of unrest occurring across the most geographically extensive state in history, offering inhabitants no political choice and no legitimate forms of political dissent. It also had access to deep, liquid foreign financial markets where it placed sovereign bonds repeatedly for more than a century. A combination of hand-collected, mapped unrest data and hand-collected, digitized foreign market bond price data allowed us to muster evidentiary support for the general effects of unrest on sovereign risk and, more importantly, the effects of the location of unrest.

Results regarding unrest location and sovereign risk generally supported the projection perspective. Importantly, our results showed that unrest events in imperial territories increased yields and spreads more than similar unrest events occurring in Russian homeland territories. Foreign bond investors demanded higher yields in the wake of unrest events taking place in territories apparently more prone to contestation and more costly to quell. While events, for the most part, did not have a direct effect on the volatility of bond yields and risk spreads, volatility was a crucial component of the pricing behavior, as the volatility created by unrest fed through to conditional means (for both yields and spreads) in nearly every single specification (including in the Heckman models). This result was robust to changes in the definition of distance and sample differentiation, and this result held across event studies of individual episodes of unrest, using never-before-digitized daily bond data, which showed how even remote rebellions affected bond yields far more than widespread unrest in St. Petersburg. One rebellion in the political heart of the Russian Empire prompted no indication of increased concern of timely sovereign bond repayment, while another rebellion on the imperial periphery surrounded by exportable resource wealth prompted a significant increase in those same concerns.

Some might respond that the projection perspective provided a better explanation of unrest location effects in the context of Tsarist Russia *only* because of distinctive historical factors in Russia's imperial expansion during the 19th and 20th centuries. Context matters and Russia's is distinctive. For another country and time, we may well find that unrest effects on sovereign risk are more consistent with the proximity perspective. Even so, our study favoring the projection perspective may provide helpful guidance for understanding how foreign creditors think about sovereign creditworthiness in geographically extensive countries dealing with the costs of remote unrest today. For example, an often-brutal armed conflict with separatists in the remote northern province of Tigray in Ethiopia has dogged the government in Addis Ababa since 2020 and seen the country's public debt increase from USD 35 billion to more than USD 55 billion over the same two years. In late 2022, the prospect of renegotiating payment terms with foreign creditors and the International Monetary Fund depends in part on whether the Ethiopian government will reverse its current military operations and negotiate a longer-term agreement ceding more local autonomy (Do Rosario and Savage, 2022). Our study of unrest location and sovereign risk in Tsarist Russia in the 19th and early 20th centuries helps us understand why governments are so willing to repress remotely located unrest but also to understand the price they pay in terms of sovereign risk and international reputation.

As noted in the introduction, political instability and unrest can both increase sovereign risk but in different ways, and thus need to be carefully separated as an analytical framework. Political instability more often decreases the *willingness* of governments to service debt while unrest more often decreases the *ability* of governments to do the same. In Tsarist Russia of the 19th and early 20th centuries, there were no elections, no parties, and no parliamentary policy debates, and unrest was the only access to the political system. However, in practically any modern country setting, both political instability and unrest can potentially affect sovereign risk. Thus, future research on unrest location and sovereign risk should control for underlying political instability individually and interactively with unrest.

Moreover, data limitations with Russia can also show if the *intensity* of unrest matters; as we noted above, we were unable to collate data on the number of fatalities of each unrest event given incomplete information and (in some instances) wide confidence bands on actual numbers. Similarly, our database is comprehensive but not exhaustive, meaning that a tally of unrest events per month across the Empire to indicate intensity was also not feasible at this juncture.

Despite these limitations, our results suggest that location matters. Perhaps most interestingly, our research also showed that Ukraine has been costly for the Russian Empire since at least the 19th century (and likely before). Of all the effects of unrest on sovereign creditworthiness uncovered in our analysis, the effect of unrest in Ukraine was the largest, with each event of unrest increasing yields by 6.9 basis points. Mark Twain told us more than 100 years ago that "[h]istory never repeats itself, but it does often rhyme." The rhyme of Ukrainian unrest then and now reminds Russia and the world of empire's high price.

**Figure 1: Unrest event locations in the Russian Empire, 1820-1914**

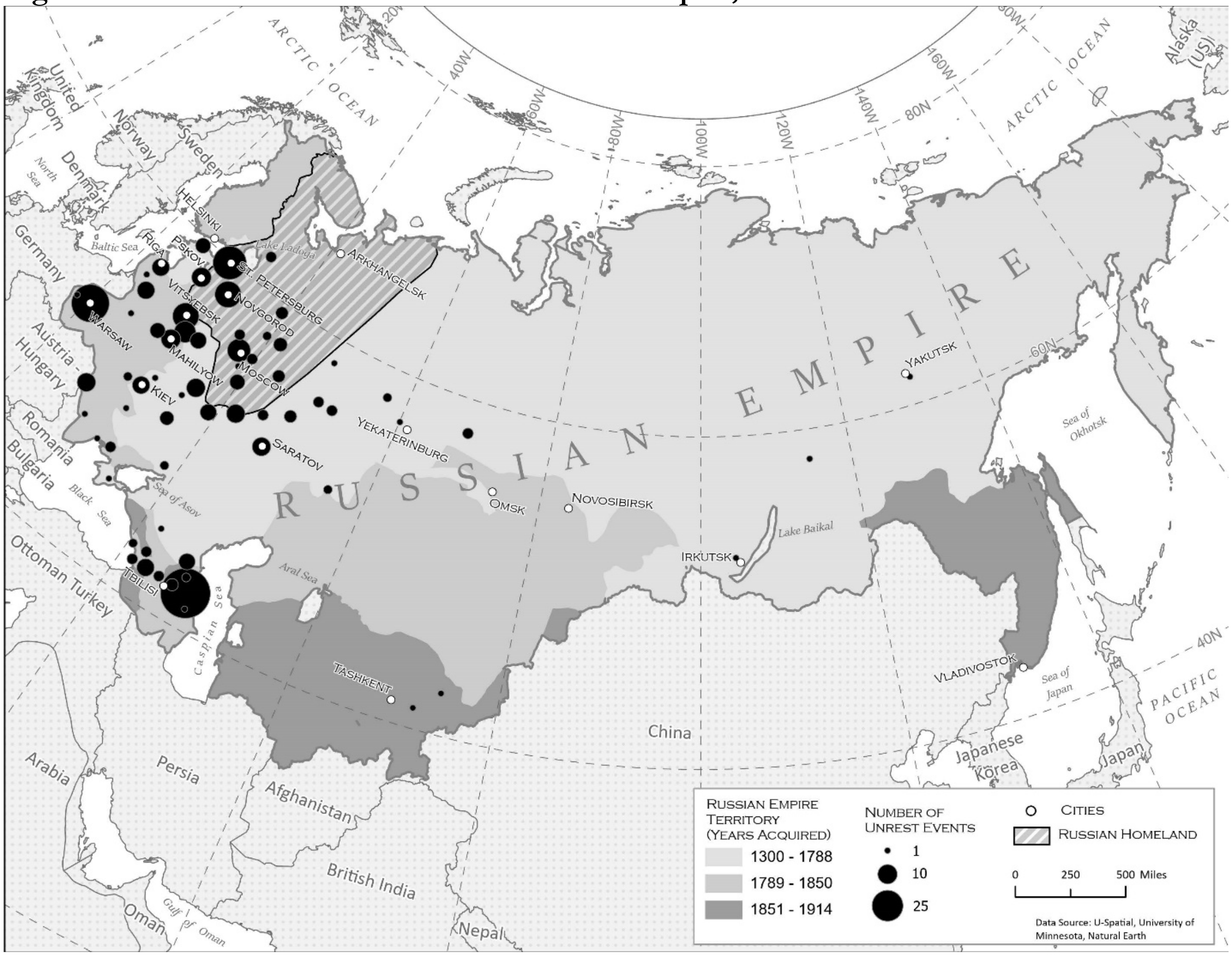

Figure 1 depicts the locations of 230 unrest events occurring in the Russian Empire from 1820-1914. Russian Empire borders and city names as well as surrounding state names and borders are current as of 1914. Larger (smaller) black dots depict locales with more (fewer) unrest events. There are 33 successful assassination attempts and 11 attempted but unsuccessful assassination attempts occurring in Russian homeland territories and 9 successful assassination attempts and 15 attempted but unsuccessful assassination attempts occurring in imperial territories. A further 85 events are collective unrest events occurring in Russian homeland territories while there are 77 of the same unrest events occurring in imperial territories. Figure 1 does not depict the additional eight external unrest events occurring at or near the borders of the Russian Empire from 1820-1914, namely wars of territorial conquest and national defense against neighboring countries or groups.

**Figure 2: Russian government bond price trends before and after the death of Tsar Alexander I on December 1, 1825**

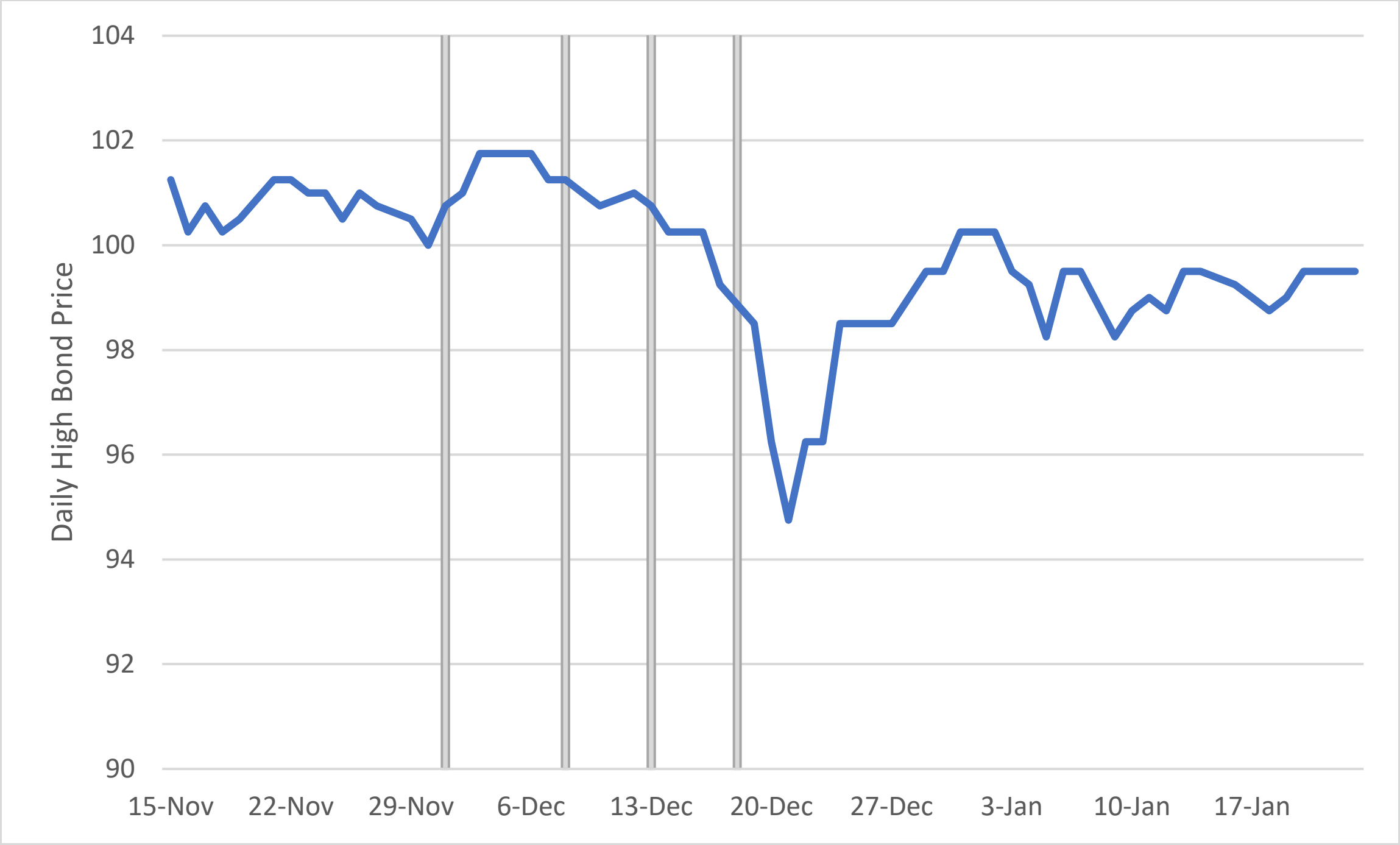

Figure 2 depicts trends in the highest price (in British pounds) each day for 5-year Russian government bonds before and after the death of Tsar Alexander I on December 1, 1825. Gray bars note four events connected with Alexander's death. The first gray bar falls on December 1, the actual date of the Alexander's sudden death from typhus while visiting the southern city of Taganrog on the Sea of Azov. The second gray bar falls on December 8, the date news of Alexander's death reached Warsaw. The third gray bar falls on December 13, the date the British government in London received news of Alexander's death. The fourth gray bar falls on December 18, the date Paris newspapers announced Alexander's death. Data for Figure 2 are from two contemporary Amsterdam broadsheets downloaded from the Capital Amsterdam Foundation (Capital Amsterdam, 2022) and digitized by the authors.

**Figure 3: Russian government bond price trends before and after the birth of Tsarevitch Alexei on August 12, 1904**

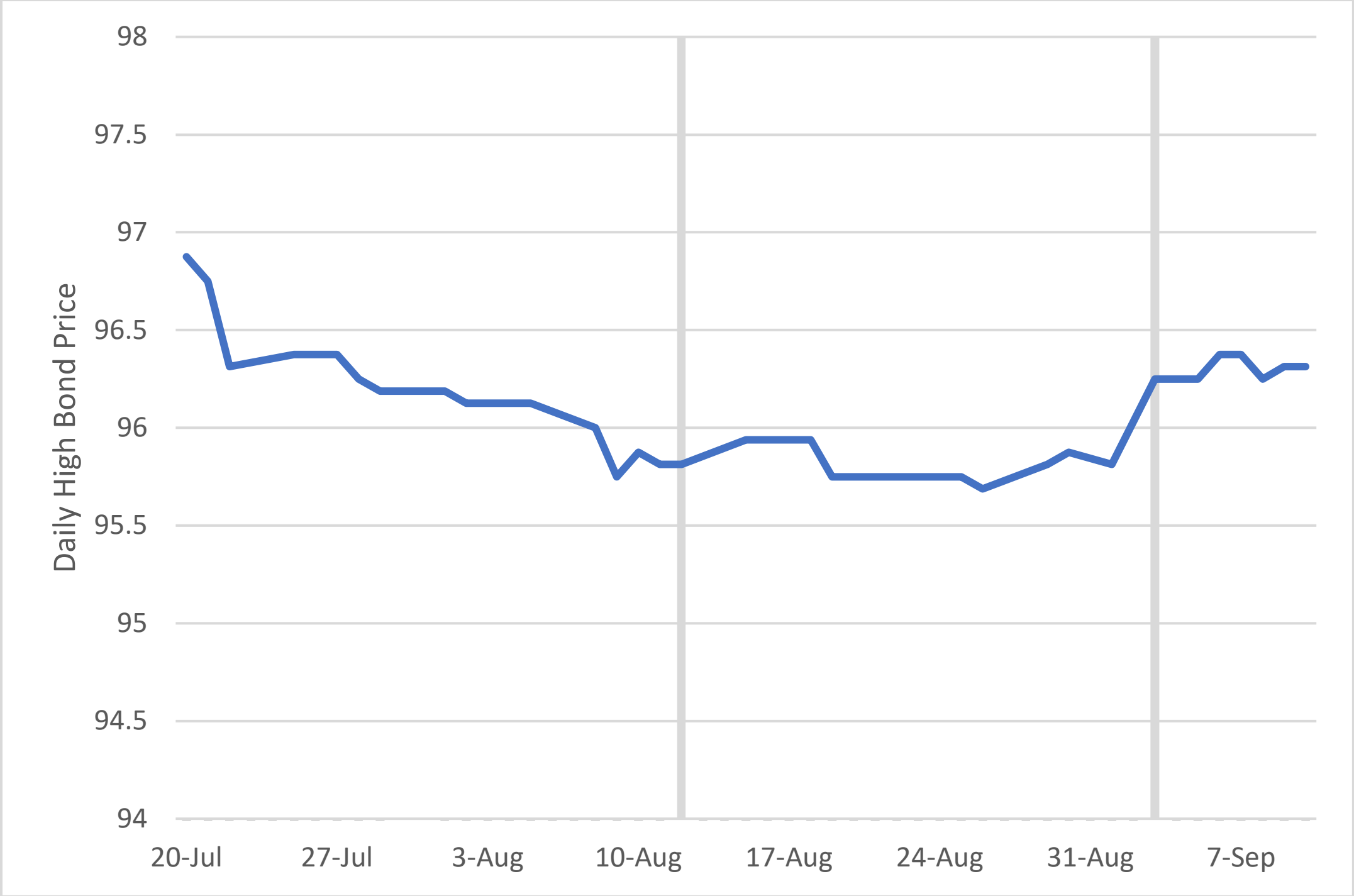

Figure 3 depicts trends in the highest price (in British pounds) each day for 5-year Russian government bonds traded in Amsterdam before and after the birth of Tsarevich Alexei on August 12, 1904. Gray bars note two events connected with Alexei's birth. The first gray bar falls on August 12, the date Alexei was born in St. Petersburg and nearly died from blood loss due to hemophilia. His parents decided to keep his hemophilia a secret from the public and the government. The second gray bar falls on September 3, the date of Alexei's Christening. During this period, news of Alexei's birth and Christening (but not his hemophilia) were communicated to and publicized almost instantaneously by Russian and foreign newspapers. Data for Figure 3 are from two contemporary Amsterdam broadsheets downloaded from Capital Amsterdam Foundation (Capital Amsterdam, 2022) and digitized by the authors.

**Figure 4: Russian government bond market liquidity trends, 1788-1914**

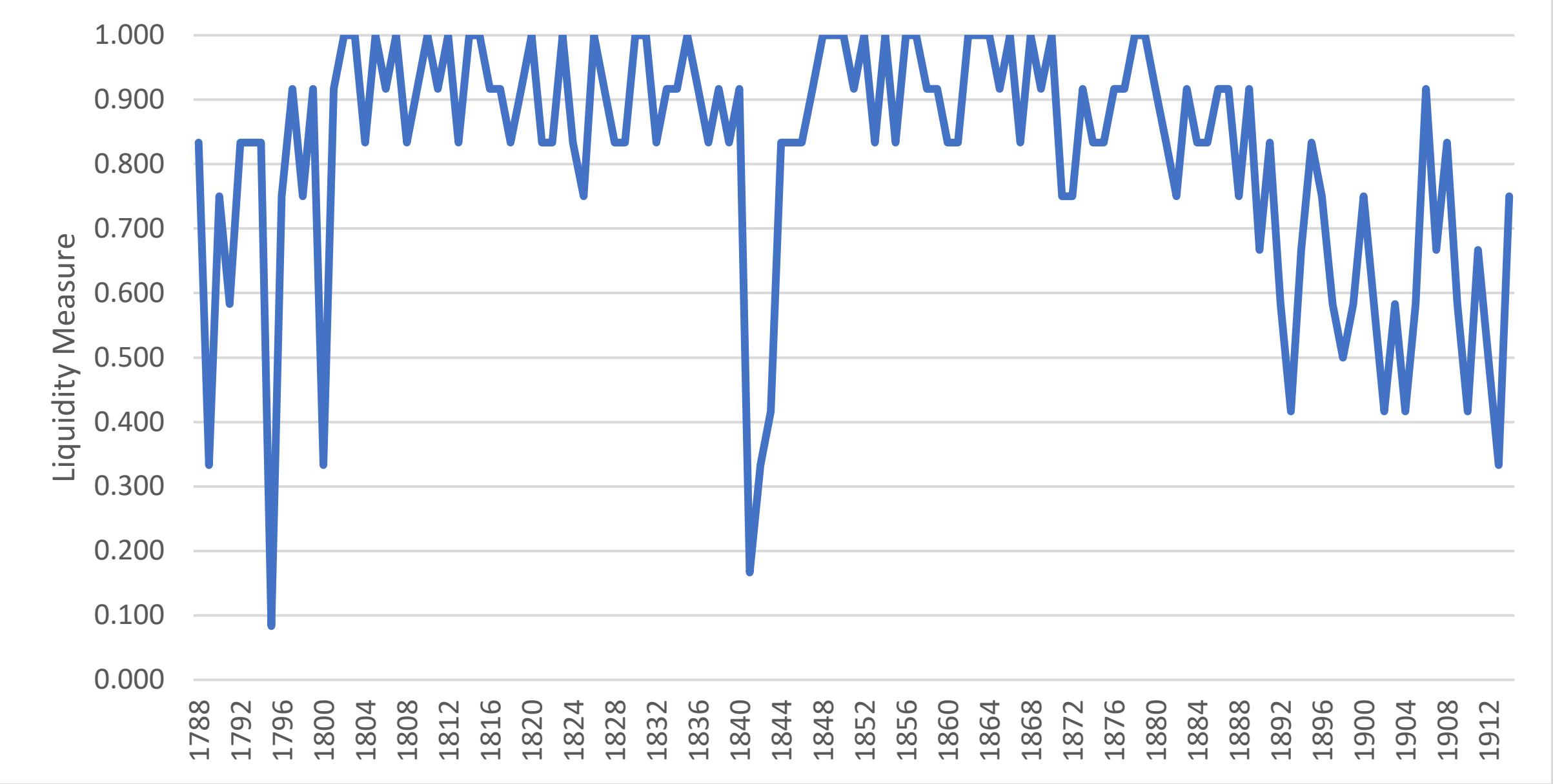

Figure 4 depicts trends in the liquidity of 5-year Russian government bonds traded in Amsterdam from 1788-1823 and then in London from 1823-1914. Years are on the x-axis and a 0-1 measure of average annual bond liquidity is on the y-axis. Average market liquidity for bond *i* in year *t* is given by Bond Liquidity = 1 – [(Number of months in year *t* that bond *i* has zero returns)/12]. Scores near 1 (0) indicate higher (lower) average annual bond market liquidity. Our measure of average annual bond market liquidity follows Campbell et al. (2018). Data for Figure 4 are from Global Financial Data (2022).

**Figure 5: Russian government bond yield and spread trends, 1788-1914**

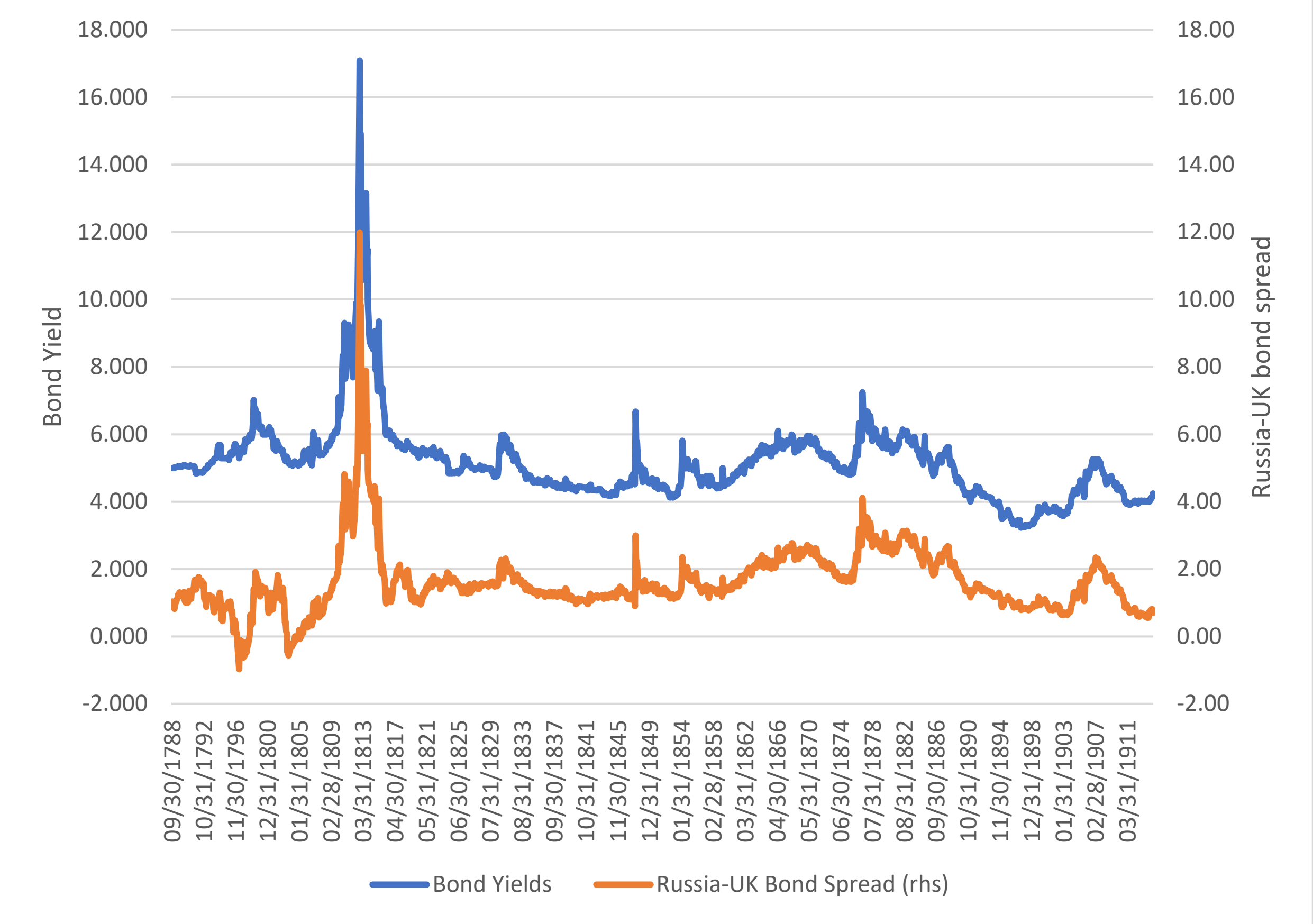

Figure 5 depicts trends in price discovery for 5-year Russian government bonds and their derivatives traded in Amsterdam from 1788-1823 and then in London from 1823-1914. Years are on the x-axis. The left-hand side of the y-axis measures yield percentages for successive 10-year Russian government bonds while the right-hand side y-axis measures spread percentages for the same Russian bonds above (+) or below (-) benchmark British government bonds. Data for Figure 5 are from Global Financial Data (2022).

**Table 1: Variable names, descriptions and data sources, descriptive statistics, and expected impact on sovereign risk, 1820-1914**

| Variable Name | Description and Data Source (Relevant Hypothesis) | Descriptive Statistics | Expected Impact on *Sovereign Risk* |
|---|---|---|---|
| *Sovereign Risk* | 5-year Russian government bond monthly percentage yields or interest-point spreads relative to benchmark British 10-year government bonds. **Source**: Global Financial Data (2022). Descriptive statistics for monthly yields and spreads presented to the right. Selected daily 5-year Russian government bond percentage yields are used in Figures 2-3 and Tables 5-6. **Source**: Capital Amsterdam (2022). Descriptive statistics for daily yields available from the authors. | Yields: Mean =4.810, Min = 3.23, SD =0.696, Max = 7.25. Spreads: Mean =1.650, Min = 0.558, SD = 0.611, Max =4.113. | Dependent Variables |
| *Attempted Assassination Empire* | 0-1 dummy equaling one (and zero otherwise) when the unrest event is an unsuccessful attempt to assassinate a specific political figure anywhere in the Russian Empire and the attempt results in fatalities. **Source**: Authors' hand-collected data. | Mean = 0.023, Min = 0, SD = 0.150, Max = 1. | Positive (Increase) |
| *Successful Assassination Empire* | 0-1 dummy equaling one (and zero otherwise) when the unrest event is a successful attempt to assassinate a specific political figure anywhere in the Russian Empire where the attempt may also result in other fatalities. **Source**: Authors' hand-collected data. | Mean = 0.035, Min = 0, SD =0.183, Max = 1. | Positive |
| *Collective Unrest Empire* | 0-1 dummy equaling one (and zero otherwise) when the unrest event is either an internal revolt, strike, demonstration, riot, mutiny, or rebellion by groups against political figures or policies anywhere in the Russian Empire resulting in fatalities. **Source**: Authors' hand-collected data. | Mean =0.214, Min = 0, SD =0.411, Max = 1. | Positive |
| *External Unrest* | 0-1 dummy equaling one (and zero otherwise) when the unrest event is either a war of territorial conquest or national defense against an external (to the Russian Empire) country or group resulting in fatalities. **Source**: Authors' hand-collected data. | Mean =0.115, Min = 0. SD =0.319, Max = 1. | Positive |
| *Cumulative Unrest Empire* | Count of cumulative unrest events in the Russian Empire over the previous 11 months from focal cumulative unrest event in the Russian Empire. **Source**: Authors' hand-collected data. | Mean =2.559, Min = 0, SD = 3.330, Max = 12.[c] | Positive |
| *Attempted Assassination Homeland* | 0-1 dummy equaling one (and zero otherwise) when the unrest event is an unsuccessful attempt to assassinate a specific political figure anywhere in the Russian homeland (See Figure 1) and the attempt results in fatalities. **Source**: Authors' hand-collected data. | Mean = 0.010, Min = 0, SD = 0.098, Max = 1. | Proximity: More Positive Projection: Less Positive |
| *Cumulative Attempted Assassinations Homeland* | Count of attempted assassination events in Russian homeland territories over the previous 11 months from focal attempted assassination event in Russian homeland territories. **Source**: Authors' hand-collected data. | Mean = 0.117, Min = 0, SD = 0.414, Max = 12. | Proximity: More Positive Projection: Less Positive |
| *Attempted Assassination Imperial* | 0-1 dummy equaling one (and zero otherwise) when the unrest event is an unsuccessful attempt to assassinate a specific political figure anywhere in imperial territories (See Figure 1) and the attempt results in fatalities. **Source**: Authors' hand-collected data. | Mean = 0.013, Min = 0, SD = 0.115, Max = 1. | Proximity: Less Positive Projection: More Positive |
| *Cumulative Attempted Assassinations Imperial* | Count of attempted assassination events in imperial territories over the previous 11 months from focal attempted assassination event in imperial territories. **Source**: Authors' hand-collected data. | Mean = 0.16, Min = 0, SD = 0.535, Max = 12. | Proximity: Less Positive Projection: More Positive |
| *Successful Assassination Homeland* | 0-1 dummy equaling one (and zero otherwise) when the unrest event is a successful attempt to assassinate a specific political figure anywhere in the Russian homeland where the attempt may also result in other fatalities. **Source**: Authors' hand-collected data. | Mean = 0.029, Min = 0, SD = 0.169, Max = 1. | Proximity: More Positive Projection: Less Positive |
| *Cumulative Successful Assassinations Homeland* | Count of successful assassination events in Russian homeland territories over the previous 11 months from focal successful assassination event in Russian homeland territories. **Source**: Authors' hand-collected data. | Mean = 0.352, Min = 0, SD = 1.176, Max = 12. | Proximity: More Positive Projection: Less Positive |
| *Successful Assassination Imperial* | 0-1 dummy equaling one (and zero otherwise) when the unrest event is a successful attempt to assassinate a specific political figure anywhere in imperial territories (See Figure 1) where the attempt may also result in other fatalities. **Source**: Authors' hand-collected data. | Mean = 0.008, Min = 0, SD =0.089, Max = 1. | Proximity: Less Positive Projection: More Positive |
| *Cumulative Successful Assassinations Imperial* | Count of successful assassination events in imperial territories over the previous 11 months from focal successful assassination event in imperial territories. **Source**: Authors' hand-collected data. | Mean = 0.096, Min = 0, SD = 0.421, Max = 12. | Proximity: Less Positive Projection: More Positive |
| *Collective Unrest Homeland* | 0-1 dummy equaling one (and zero otherwise) when the unrest event is either an internal revolt, strike, demonstration, riot, mutiny, or rebellion by groups against political figures or policies anywhere in the Russian homeland (See Figure 1) resulting in fatalities. **Source**: Authors' hand-collected data. | Mean =0.143, Min = 0, SD = 0.350, Max = 1. | Proximity: More Positive Projection: Less Positive |
| *Cumulative Collective Unrest Homeland* | Count of collective unrest events in Russian homeland territories over the previous 11 months from collective unrest event in Russian homeland territories. **Source**: Authors' hand-collected data. | Mean = 1.589, Min = 0, SD = 2.795, Max = 12. | Proximity: More Positive Projection: Less Positive |
| *Collective Unrest Imperial* | 0-1 dummy equaling one (and zero otherwise) when the unrest event is either an internal revolt, strike, demonstration, riot, mutiny, or rebellion by groups against political figures or policies in imperial territories (See Figure 1) resulting in fatalities. **Source**: Authors' hand-collected data. | Mean =0.190, Min = 0, SD =0.393, Max = 1. | Proximity: Less Positive Projection: More Positive |
| *Cumulative Collective Unrest Imperial* | Count of collective unrest events in imperial territories over the previous 11 months from focal collective unrest event in imperial territories. **Source**: Authors' hand-collected data. | Mean = 1.812, Min = 0, SD = 2.912, Max = 12. | Proximity: Less Positive Projection: More Positive |
| *Collective Unrest Imperial plus Caucasus Rebellion* | 0-1 dummy equaling one (and zero otherwise) when the unrest event is either an internal revolt, strike, demonstration, riot, mutiny, or rebellion by groups against political figures or policies in imperial territories (See Figure 1) resulting in fatalities. This also includes the aftermath of the Caucasus campaigns and repeated rebellions until the Caucasus were pacified in the 1860s. **Source**: Authors' hand-collected data. | Mean =0.156, Min = 0, SD =0.363, Max = 1. | Proximity: Less Positive Projection: More Positive |
| *Collective Unrest Imperial plus Caucasus Rebellion and Caucasus Wars* | 0-1 dummy equaling one (and zero otherwise) when the unrest event is either an internal revolt, strike, demonstration, riot, mutiny, or rebellion by groups against political figures or policies in imperial territories (See Figure 1) resulting in fatalities. This includes every event from the previous two Collective Unrest variables but also classifies the Caucasus campaign itself (various dates from 1817 to 1864) as political instability rather than a war of conquest. **Source**: Authors' hand-collected data. | Mean =0.188, Min = 0, SD =0.391, Max = 1. | Proximity: Less Positive Projection: More Positive |
| *Collective Unrest Ukraine* | 0-1 dummy equaling one (and zero otherwise) when a collective unrest event occurs in imperial territories comprising modern-day Ukraine. **Source**: Authors' hand-collected data. | Mean = 0.417, Min = 0, SD = 0.200, Max = 1. | Proximity: Less Positive Projection: More Positive |
| *Tsar Transition* | 0-1 dummy equaling one (and zero otherwise) when there is an unrest event falling in the same month *t* as a change in tsars. **Source**: Lester (1995) | Mean = 0.004, Min = 0, SD = 0.067, Max = 1. | Positive |
| *Gold* | Monthly change in the price of gold bullion (London) in month *t*. **Source**: Global Financial Data (2022). | Mean = 0.001, Min = -10.439, SD = 0.714, Max = 11.363. | Negative (Decrease) |
| *Ruble-Guilder* | Monthly percentage of Russian ruble-Dutch guilder exchange in month *t*. **Source**: Global Financial Data (2022). | Mean = -0.0003, Min = -0.154, SD = 0.019, Max = 0.094. | Negative |
| *Drought* | Annual (June-August) 0-3 measure of Palmer Drought Severity Index (PDSI) on a 0.5° latitude/longitude grid running from Russia's East European Plain to the Ural Mountains. Higher PDSI values indicate more intense drought conditions during the growing season. **Source**: Cook et al. (2020). | Mean = 0.128, Min = 0 SD = 0.444, Max = 3. | Positive |
| *Distance* | The natural log of kilometers from unrest event location anywhere in the Russian Empire to St. Petersburg. **Source**: Authors' hand-collected data[a] | Mean = 5.842, Min = -2.303 SD = 2.511, Max = 8.393 | Proximity: Negative Projection: Positive |
| *Serfdom* | 0-1 dummy equaling one (and zero otherwise) if the unrest event occurred prior to publication of Tsar Alexander II's Emancipation Manifesto in March 1861. | Mean = 0.576, Min = 0, SD = 0.464, Max = 1. | Negative (Both Stages) |
| *Interior* | Percent of total annual Russian imperial court budget allocated the Ministry of the Interior in month *t*, smoothed using Baxter-King (1999) filter. **Source**: Borodkin (2022a). | Mean = 0.000, Min = -0.030, SD = 0.005, Max = 0.028. | Negative (Both Stages) |
| *Cereals* | Natural log of annual ruble value of all cereal exports from the Russian Empire in month *t*. **Source**: Borodkin (2022b). | Mean = 11.290, Min = 8.403. SD = 1.358, Max = 13.528. | Negative (Both Stages) |
| *Size* | Square kilometers of oblast (region) in month *t*. **Source**: Lahmeyer (2006). | Mean = 65189, Min = 11, SD = 161396, Max = 3103200. | Positive (Distance) |
| *Density* | Number of people per square kilometer in oblast (region) in month *t*. **Sources**: Lahmeyer (2006) and Lordkipanidze and Totadze (2010). | Mean = 77.558, Min = 0.030, SD = 221.412, Max = 1611.258. | Negative (Distance) |
| *Lost War* | 0-1 dummy equaling one (and zero otherwise) in the month following a loss of a war (practically, following the Crimean and Russo-Japanese Wars). | Mean = 0.043, Min = 0, SD = 0.203, Max = 1. | Positive (Distance) |
| *Bond Volatility* | Realized volatility of yield or risk spreads, calculated as $\ln\left({Price_t}/{Price_{t-1}}\right)^2$ of the particular financial metric. | Mean = 0.0009, Min = 0, SD = 0.005, Max = 0.153 | Negative |

a. For example, Russian royal cartographers in the 19th century determined that the distance from St. Petersburg to Warsaw was 852.2 versts, a unit of distance no longer used. One verst is approximately 1.07 kilometers. Thus, the distance in kilometers was 909.13. Most verst-based distance measures were taken from the *Geographical Alas of the Russian Empire, the Kingdom of Poland, and the Grand Duchy of Finland*, produced by the Military-Topographical Depot of the Russian General Staff (Atlas, 2022). Other distance measures not covered in the atlas derive from contemporary Google-based interactive maps following historical roadways.

**Figure 6: Mean sovereign risk measures for unrest event types in the Russian Empire, 1820-1914**

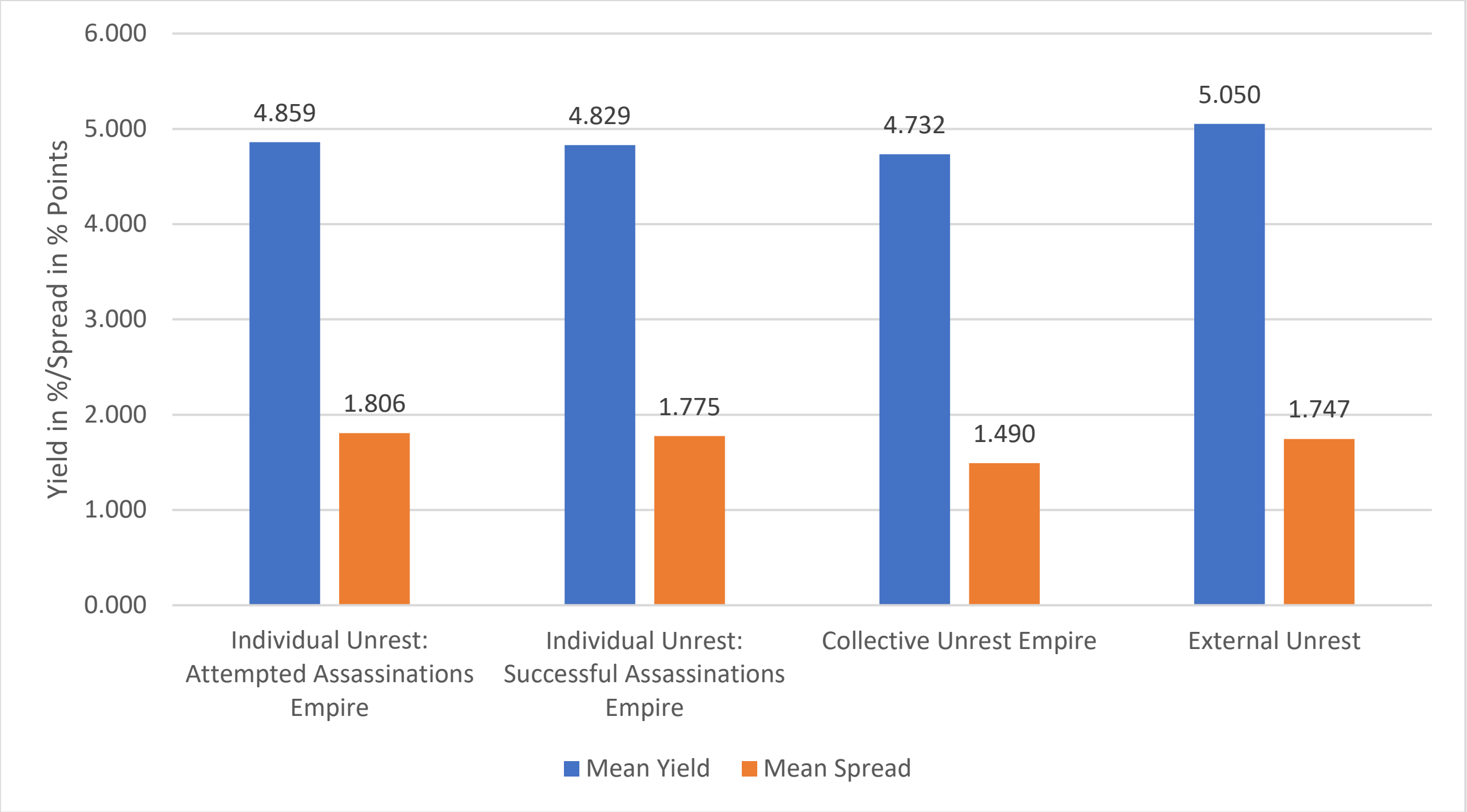

Figure 6 depicts mean 5-year Russian government bond yield percentages and spread percentage points compared to benchmark 10-year British government bonds for four different types of unrest events occurring anywhere in the Russian Empire from 1820-1914: Individual unrest in the form of attempted and successful assassinations; collective unrest events; and external unrest events. Each unrest event type is defined in Table 1. Data for Figure 6 on different unrest event types are hand-collected by the authors. Data for Figure 6 on mean 5-year Russian bond yields and spreads come from Global Financial Data (2022).

**Figure 7: Mean sovereign risk measures for unrest event types in Russian homeland and imperial territories, 1820-1914**

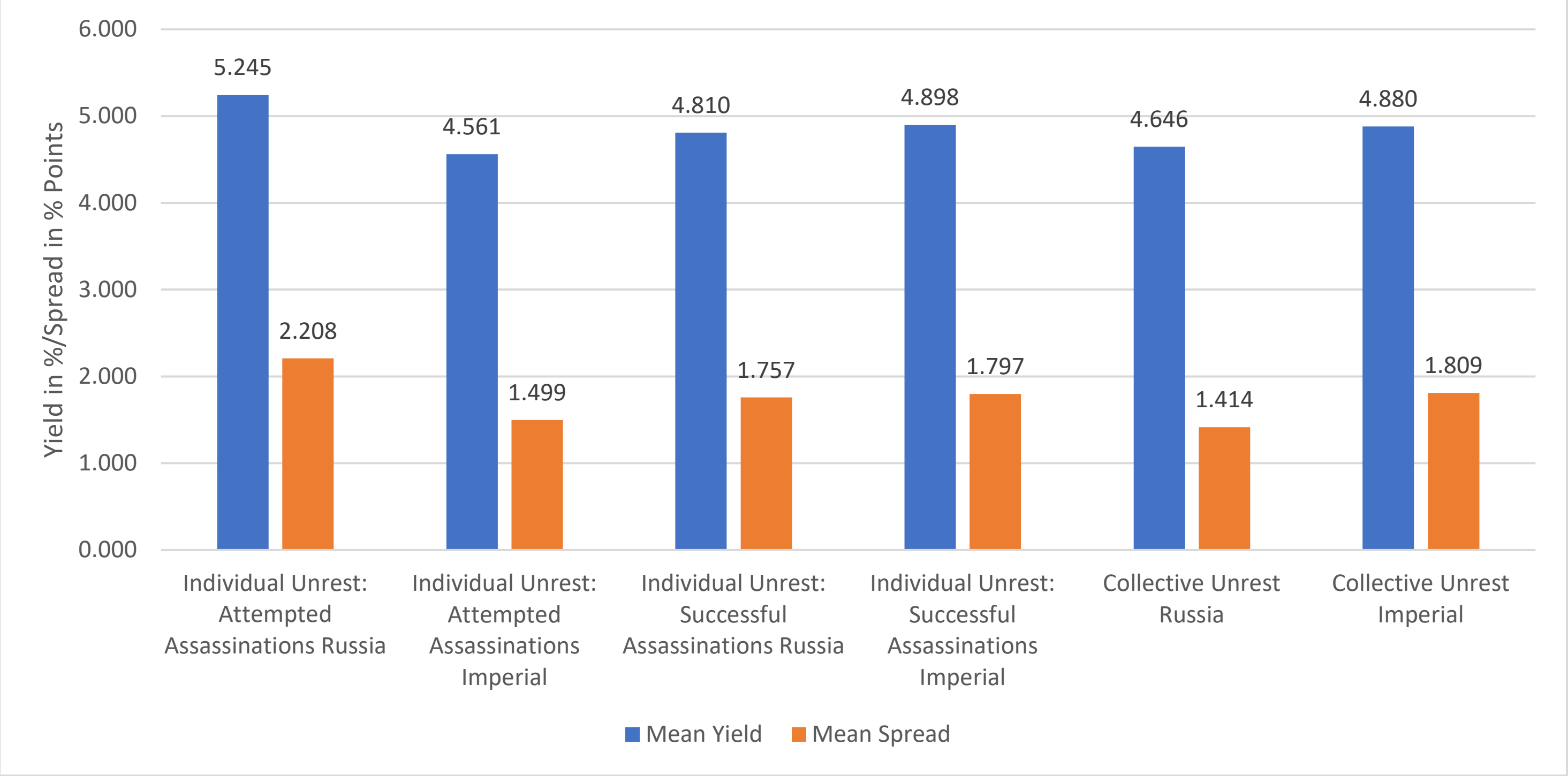

Figure 7 depicts mean 5-year Russian government bond yield percentages and spread percentage points compared to benchmark 10-year British government bonds for four different types of unrest events occurring Russian homeland and imperial territories from 1820-1914: individual unrest in the form of attempted and successful assassinations; collective unrest events; and external unrest events. Russian homeland and imperial territories are depicted in Figure 1. Each unrest event type is defined in Table 1. Data for Figure 7 on different unrest event types are hand-collected by the authors. Data for Figure 7 on mean 5-year Russian bond yields and spreads come from Global Financial Data (2022).

**Figure 8: Average Abnormal Returns Surrounding Collective Unrest in Ukraine**

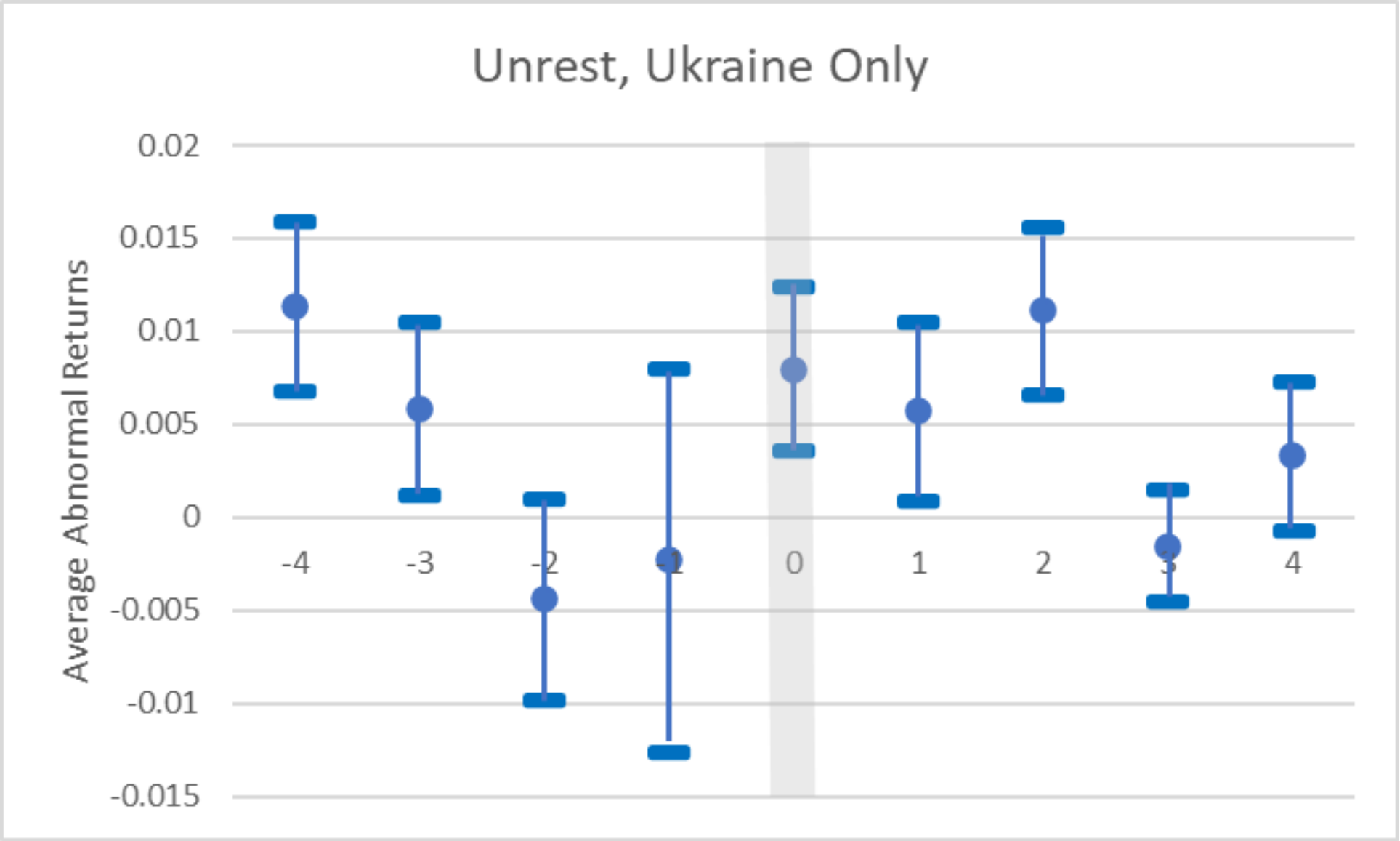

Figure 8 depicts average abnormal returns to bond yields over a four-month window before and after a collective unrest event occurring in the imperial territories of modern-day Ukraine. The top and bottom bars represent the 95% confidence interval for the abnormal returns using the Adjusted Patell test, while the circle represents the estimated mean.

**Table 2: Results from Raw Returns Event Study Regressions of single unrest events and sovereign risk in Russian Empire, 1820-1914**

| Raw Returns Model | | | |
|---|---|---|---|
| **Bond Returns** | **n** | [-1,+1] **CAAR** | |
| | | **Patell Adjusted** | **GRANKT** |
| Attempted Assassination Empire | 26 | 0.006 | |
| | | *1.15* | *1.12* |
| Attempted Assassination Homeland | 11 | -0.013 | |
| | | *2.21*** | *2.90**** |
| Attempted Assassination Imperial | 15 | 0.020 | |
| | | *3.21**** | *2.72**** |
| Successful Assassination Empire | 39 | 0.010 | |
| | | *3.27**** | *1.76** |
| Successful Assassination Homeland | 33 | 0.011 | |
| | | *3.02**** | *1.23* |
| Successful Assassination Imperial | 9 | 0.018 | |
| | | *2.96**** | *2.34*** |
| Unrest Empire | 240 | 0.0005 | |
| | | *4.21**** | *0.55* |
| Unrest Homeland | 160 | 0.003 | |
| | | *2.63**** | *0.46* |
| Unrest Imperial | 141 | 0.011 | |
| | | *5.84**** | *3.07**** |
| Unrest Imperial with Caucasus Events | 189 | 0.010 | |
| | | *6.48**** | *4.74**** |
| Unrest Imperial with Caucasus Events and War | 197 | 0.010 | |
| | | *6.32**** | *4.37**** |
| External Conflict | 129 | 0.004 | |
| | | *1.73** | *1.25* |

Table 2 reports results from event study analyses of abnormal 5-year Russian government bond yield percentage returns (yield returns) for each category of political unrest. Event study output includes yield returns for one month prior and one month following the event. Model run using raw returns as baseline for calculating abnormal returns. Patell Adjusted is the parametric test of Kolari and Pynnönen (2010) while GRANKT is the generalized rank t-test of Kolari and Pynnönen (2011). Absolute values of t-stats shown in italics at statistical significance levels: *** = $p < 0.01$ (1% level), ** = $p < 0.05$ (5% level), * = $p < 0.10$ (10% level).

**Table 3: Results from Constant Mean Event Study Regressions of single unrest events and sovereign risk in Russian Empire, 1820-1914**

| Constant Mean Model | | | |
|---|---|---|---|
| **Bond Returns** | **n** | [-1,+1] | |
| | | **CAAR** | |
| | | **Patell Adjusted** | **GRANKT** |
| Attempted Assassination Empire | 26 | 0.007 | |
| | | *1.46* | *0.93* |
| Attempted Assassination Homeland | 11 | -0.013 | |
| | | *2.21*** | *3.10**** |
| Attempted Assassination Imperial | 15 | 0.023 | |
| | | *3.65**** | *2.88**** |
| Successful Assassination Empire | 39 | 0.013 | |
| | | *3.96**** | *2.16*** |
| Successful Assassination Homeland | 33 | 0.014 | |
| | | *3.69**** | *1.57* |
| Successful Assassination Imperial | 9 | 0.020 | |
| | | *3.33**** | *2.55*** |
| Unrest Empire | 240 | 0.007 | |
| | | *5.88**** | *3.94**** |
| Unrest Homeland | 160 | 0.005 | |
| | | *3.91**** | *1.94** |
| Unrest Imperial | 141 | 0.012 | |
| | | *6.85**** | *5.27**** |
| Unrest Imperial with Caucasus Events | 189 | 0.012 | |
| | | *7.64**** | *6.75**** |
| Unrest Imperial with Caucasus Events and War | 197 | 0.011 | |
| | | *7.43**** | *6.20**** |
| External Conflict | 129 | 0.007 | |
| | | *3.73**** | *5.65**** |

Table 3 reports results from event study analyses of abnormal 5-year Russian government bond yield percentage returns (yield returns) for each category of political unrest. Event study output includes yield returns for one month prior and one month following the event. Model run using constant mean returns as baseline for calculating abnormal returns. Patell Adjusted is the parametric test of Kolari and Pynnönen (2010) while GRANKT is the generalized rank t-test of Kolari and Pynnönen (2011). Absolute values of t-stats shown in italics at statistical significance levels: *** = p < 0.01 (1% level), ** = p < 0.05 (5% level), * = p < 0.10 (10% level).

**Table 4: Results from event study of abnormal Russian government bond returns surrounding the Obukhov Defense in St. Petersburg, May 1901**

| Start of Strike, May 1, 1901 | | | | | |
|---|---|---|---|---|---|
| **Event Windows →** <br> ↓*Estimates* | **1** <br> **[-1,1]** | | **2** <br> **[-3,3]** | | **3** <br> **[-5,5]** |
| *Historical average window closed at t =-30.* | | | | | |
| Percentage Russian Government Bond Returns | -0.82* | | -0.84 | | -0.87 |
| Volatility | 0.00 | | -0.001 | | -0.001 |
| Brownian Intraday Volatility | -0.02 | | -0.14 | | -0.25 |
| *Historical average window closed at t =-20.* | | | | | |
| Percentage Russian Government Bond Returns | -0.84* | | -0.89 | | -0.94 |
| Volatility | 0.00 | | -0.001 | | -0.001 |
| Brownian Intraday Volatility | -0.03 | | -0.15 | | -0.28 |
| **Worker Clashes with Police and Army, May 7, 1901** | | | | | |
| **Event Windows →** <br> ↓*Estimates* | **1** <br> **[-1,1]** | | **2** <br> **[-3,3]** | | **3** <br> **[-5,5]** |
| *Historical average window closed at t =-30.* | | | | | |
| Percentage Russian Government Bond Returns | -0.03 | | -0.07 | | -0.36 |
| Volatility | -0.001 | | -0.001 | | -0.001 |
| Brownian Intraday Volatility | -0.12 | | -0.24 | | -0.27 |
| *Historical average window closed at t =-20.* | | | | | |
| Percentage Russian Government Bond Returns | -0.03 | | -0.06 | | -0.36 |
| Volatility | -0.001 | | -0.001 | | -0.010 |
| Brownian Intraday Volatility | -0.13 | | -0.25 | | -0.28 |

Table 4 reports results from event study analyses of abnormal 5-year Russian government bond yield percentage returns (yield returns) during unrest associated with the Obukhov Defense in St. Petersburg during May 1901. Event study output includes yield returns and two related volatility estimates. This event study output is reported for two event dates (May 1, 1901; May 7, 1901), with two pre-event historical average yield windows (t = -30; t = -20), and three event windows (-1, 1; -3, 3; -5, 5). Statistical significance levels include: *** = p < 0.01 (1% level), ** = p < 0.05 (5% level), * = p < 0.10 (10% level).

**Table 5: Results from event study analysis of abnormal Russian government bond returns surrounding the arrest of Ivan Kovalsky in Odesa, February 11, 1878**

| **Event Windows →**<br>↓*Estimates* | **1**<br>**[-1,1]** | | **2**<br>**[-3,3]** | | **3**<br>**[-5,5]** |
|---|---|---|---|---|---|
| *Historical average window closed at t =-30.* | | | | | |
| Percentage Russian Government Bond Returns | -2.99*** | | -1.34 | | -2.29 |
| Volatility | 0.04 | | 0.06 | | 0.05 |
| Brownian Intraday Volatility | 0.68*** | | 1.66*** | | 2.01*** |
| *Historical average window closed at t =-20.* | | | | | |
| Percentage Russian Government Bond Returns | -3.01*** | | -1.37 | | -2.34 |
| Volatility | 0.04 | | 0.06 | | 0.05 |
| Brownian Intraday Volatility | 0.68*** | | 1.65*** | | 1.99*** |

Table 5 reports results from event study analyses of abnormal 5-year Russian government bond yield percentage returns (yield returns) during unrest associated with the arrest of Ivan Kovalsky in Odesa on February 11, 1878. Event study output includes yield returns and two related volatility estimates. This event study output is reported for one event date (November 21, 1893), with two pre-event historical average yield windows (t = -30, t = -20), and three event windows (-1, 1; -3, 3; -5, 5). Statistical significance levels include: *** = p < 0.01 (1% level), ** = p < 0.05 (5% level), * = p < 0.10 (10% level).

## Table 6: Results from ACGARCH-M analysis of single unrest events and sovereign risk in the Russian Empire, 1820-1914

| **Dependent Variables →** <br> *↓Independent Variables* | **1** <br> **Yields** | **2** <br> **Spreads** | **3** <br> **Yields** | **4** <br> **Spreads** | **5** <br> **Yields** | **6** <br> **Spreads** | **7** <br> **Yields** | **8** <br> **Spreads** | **9** <br> **Yields** | **10** <br> **Spreads** |
|---|---|---|---|---|---|---|---|---|---|---|
| *Attempted Assassination Empire* | | | 0.030* <br> (1.88) | 0.050*** <br> (2.66) | | | | | | |
| *Successful Assassination Empire* | | | | | -0.012 <br> (1.27) | -0.013** <br> (2.03) | | | | |
| *Collective Unrest Empire* | | | | | | | 0.004*** <br> (3.22) | 0.005*** <br> (4.82) | | |
| *External Unrest Empire* | | | | | | | | | 0.010*** <br> (3.98) | 0.026*** <br> (2.91) |
| *Tsar Transition* | 0.057** <br> (2.43) | 0.028 <br> (0.78) | 0.085** <br> (2.04) | 0.024 <br> (0.66) | 0.089** <br> (1.98) | 0.076* <br> (1.71) | 0.086* <br> (1.82) | 0.107*** <br> (4.25) | 0.081* <br> (1.90) | 0.061* <br> (1.67) |
| *Gold* | -0.008*** <br> (7.12) | -0.007** <br> (2.18) | -0.008** <br> (2.39) | -0.008** <br> (2.37) | -0.007** <br> (2.17) | -0.005 <br> (1.10) | -0.008** <br> (2.45) | -0.009** <br> (2.09) | -0.008** <br> (2.44) | -0.009** <br> (2.37) |
| *Ruble-Guilder* | -0.163** <br> (2.29) | -0.992*** <br> (5.98) | -0.531*** <br> (4.53) | -0.958*** <br> (5.66) | -0.556*** <br> (4.79) | -0.507*** <br> (2.92) | -0.586*** <br> (4.55) | -0.539*** <br> (3.56) | -0.499*** <br> (4.48) | -1.079*** <br> (5.73) |
| *Drought* | 0.001 <br> (0.31) | 0.001 <br> (0.17) | -0.003 <br> (0.57) | 0.002 <br> (0.26) | -0.003 <br> (0.56) | -0.008 <br> (1.21) | -0.004 <br> (0.84) | -0.006 <br> (1.01) | -0.004 <br> (0.77) | 0.003 <br> (0.43) |
| *Lagged Dependent Variable* | 0.999*** <br> (4691.65) | 1.003*** <br> (189.69) | 0.992*** <br> (7150.37) | 1.001*** <br> (247.26) | 0.992*** <br> (7903.84) | 0.967*** <br> (517.50) | 0.990*** <br> (1537.30) | 0.985*** <br> (486.32) | 0.987*** <br> (544.40) | 1.012*** <br> (118.61) |
| Constant | 0.005** <br> (4.51) | 0.080** <br> (2.16) | 0.033*** <br> (13.95) | -0.075*** <br> (4.36) | 0.032*** <br> (13.83) | 0.003 <br> (1.12) | 0.045*** <br> (31.76) | 0.016*** <br> (4.79) | 0.049*** <br> (7.04) | -0.169*** <br> (2.87) |
| *↓GARCH Attributes* | | | | | | | | | | |
| Unrest Event Long-Term Volatility Empire | | | -0.005 <br> (1.03) | 0.003 <br> (0.96) | 0.002 <br> (0.89) | 0.028 <br> (1.34) | -0.0003 <br> (1.50) | 0.012 <br> (0.64) | -0.0004 <br> (0.71) | 0.002 <br> (0.63) |
| Unrest Event Short-Term Volatility Empire | | | 0.012* <br> (1.89) | 0.001 <br> (0.04) | -0.003 <br> (1.27) | -0.031 <br> (1.47) | 0.001 <br> (0.89) | -0.012 <br> (0.67) | -0.002*** <br> (2.76) | -0.003 <br> (0.40) |
| GARCH-in-Mean | -0.404*** <br> (4.26) | -0.015** <br> (2.32) | -0.271** <br> (2.38) | -0.014*** <br> (4.78) | -0.358** <br> (2.10) | 0.449*** <br> (8.12) | -0.651*** <br> (3.12) | N/A | N/A | -0.030** <br> (0.010) |
| Adjusted $R^2$ | 0.963 | 0.951 | 0.964 | 0.951 | 0.963 | 0.955 | 0.964 | 0.951 | 0.962 | 0.954 |
| AIC | -2.113 | -1.913 | -2.087 | -1.916 | -2.082 | -1.916 | -2.069 | -1.867 | -2.091 | -1.909 |
| n | 1125 | 1125 | 1125 | 1125 | 1125 | 1125 | 1125 | 1125 | 1125 | 1125 |

Table 6 reports results from asymmetric component GARCH-in-mean (ACGARCH-M) analyses of single unrest event effects on sovereign risk no matter where unrest events may occur in the Russian Empire from 1820-1914. There are four unrest event types measured as 0-1 dummies taking the value of one when occurring in month *t*: individual unrest events in the forms of attempted assassinations and successful assassinations; collective unrest events of different forms (e.g., worker strikes); and external unrest events of different forms (e.g., wars of territorial conquest). We regress two measures of sovereign risk on these unrest indicators: month *t* 5-year Russian government bond yield percentages (yields) and month *t* 5-year Russian government bond spread percentage points above the benchmark 10-year British government bond yield percentage (spreads). Table 1 provides details on all variables included in these ACGARCH-M analyses. We report coefficient and volatility estimates along with absolute t-statistics (in parentheses) and levels of statistical significance for all right-hand side variables analyzed. Statistical significance levels include: *** = $p < 0.01$ (1% level), ** = $p < 0.05$ (5% level), * = $p < 0.10$ (10% level). Columns 1-2 do not include unrest event terms and, thus, long- and short-term volatility terms for unrest are neither included nor reported. GARCH-in-mean terms are dropped in Columns 8-9.

## Table 7: Results from ACGARCH-M analysis of cumulative unrest and sovereign risk in the Russian Empire, 1820-1914

| Dependent Variables → | 1 | 2 | 3 | 4 | 5 | 6 | 7 | 8 |
|---|---|---|---|---|---|---|---|---|
| ↓*Independent Variables* | **Yields** | **Spreads** | **Yields** | **Spreads** | **Yields** | **Spreads** | **Yields** | **Spreads** |
| *Cumulative Attempted Assassination Empire* | 0.008**<br>(2.16) | 0.009***<br>(2.59) | | | | | | |
| *Cumulative Successful Assassination Empire* | | | 0.012***<br>(4.10) | 0.004***<br>(2.88) | | | | |
| *Cumulative Collective Unrest Empire* | | | | | 0.009***<br>(5.31) | 0.002***<br>(3.17) | | |
| *Cumulative External Unrest Empire* | | | | | | | 0.001***<br>(2.62) | 0.003***<br>(2.93) |
| *Tsar Transition* | 0.086*<br>(1.76) | 0.082<br>(1.25) | 0.069<br>(1.07) | 0.070**<br>(2.28) | 0.164<br>(1.15) | 0.077**<br>(2.13) | 0.072*<br>(1.72) | 0.097<br>(1.13) |
| *Gold* | -0.008**<br>(2.19) | -0.008*<br>(1.89) | -0.016***<br>(3.95) | -0.008***<br>(2.74) | 0.013<br>(1.14) | -0.008**<br>(2.16) | -0.008*<br>(2.24) | -0.009**<br>(2.41) |
| *Ruble-Guilder* | -0.543***<br>(4.48) | -0.728***<br>(4.81) | -1.194***<br>(10.89) | -0.785***<br>(5.22) | -1.017***<br>(5.27) | -0.808***<br>(5.15) | -0.635***<br>(10.51) | -0.816***<br>(5.18) |
| *Drought* | -0.005<br>(0.76) | -0.002<br>(0.30) | 0.010<br>(0.87) | -0.003<br>(0.41) | 0.053***<br>(5.45) | 0.008<br>(1.18) | -0.001<br>(0.22) | 0.001<br>(0.17) |
| *Lagged DV* | 0.992***<br>(4913.63) | 0.995***<br>(346.55) | 0.991***<br>(1995.14) | 0.989***<br>(407.94) | 0.948***<br>(126.04) | 1.005***<br>(146.76) | 0.991***<br>(3938.11) | 1.001***<br>(306.59) |
| Constant | 0.035***<br>(13.35) | 0.013***<br>(2.84) | -0.388***<br>(33.36) | -0.059***<br>(3.85) | -0.353***<br>(9.54) | -0.118***<br>(2.94) | 0.037***<br>(40.14) | 0.063***<br>(4.15) |
| ↓*GARCH Attributes* | | | | | | | | |
| Unrest Event Long-Term Volatility Empire | 0.0004<br>(1.45) | -0.006<br>(1.13) | -0.002<br>(0.93) | 0.02<br>(1.45) | 0.0001<br>(0.17) | 0.004*<br>(1.76) | 0.0001<br>(0.19) | 0.024<br>(0.10) |
| Unrest Event Short-Term Volatility Empire | 0.0002<br>(0.28) | 0.007<br>(1.20) | 0.004<br>(1.47) | -0.010<br>(1.06) | 0.0003<br>(0.27) | -0.003<br>(1.32) | -0.0002<br>(0.40) | -0.024<br>(0.10) |
| GARCH-in-Mean | -0.486**<br>(2.54) | -1.484***<br>(2.72) | -0.090***<br>(34.31) | -0.472***<br>(3.35) | -0.109***<br>(44.46) | 0.020***<br>(3.24) | -0.835***<br>(4.54) | -0.775***<br>(4.81) |
| Adjusted $R^2$ | 0.964 | 0.954 | 0.963 | 0.953 | 0.829 | 0.951 | 0.964 | 0.954 |
| AIC | -2.087 | -1.899 | -1.984 | -1.853 | -2.139 | -1.877 | -2.071 | -1.889 |
| n | 1125 | 1125 | 1125 | 1125 | 1125 | 1125 | 1125 | 1125 |

Table 7 reports results from asymmetric component GARCH-in-mean (ACGARCH-M) analyses of cumulative unrest event effects on sovereign risk no matter where unrest events may occur in the Russian Empire from 1820-1914. There are four focal unrest event types measured as 0-1 dummies taking the value of one when occurring in month *t*: individual unrest events in the forms of attempted assassinations and successful assassinations; collective unrest events of different forms (e.g., worker strikes); and external unrest events of different forms (e.g., wars of territorial conquest). When these unrest event dummies take the value of one, we count the number of months in the previous year when the same type of unrest occurred anywhere in the Russian Empire. We regress two measures of sovereign risk on these cumulative unrest indicators: month *t* 5-year Russian government bond yield percentages (yields) and month *t* 5-year Russian government bond spread percentage points above the benchmark 10-year British government bond yield percentage (spreads). Table 1 provides details on all variables included in these ACGARCH-M analyses. We report coefficient and volatility estimates along with absolute t-statistics (in parentheses) and levels of statistical significance for all right-hand side variables analyzed. Statistical significance levels include: ** = $p < 0.01$ (1% level), * = $p < 0.05$ (5% level), † = $p < 0.10$ (10% level).

## Table 8: Results from ACGARCH-M analysis of single unrest events, cumulative unrest, and sovereign risk in Russian homeland and imperial territories, 1820-1914

| Dependent Variables → <br> ↓*Independent Variables* | 1 <br> Yields | 2 <br> Spreads | 3 <br> Yields | 4 <br> Spreads | 5 <br> Yields | 6 <br> Spreads | 7 <br> Yields | 8 <br> Spreads | 9 <br> Yields | 10 <br> Spreads | 11 <br> Yields | 12 <br> Spreads |
|---|---|---|---|---|---|---|---|---|---|---|---|---|
| | Single Events | | | | | | Cumulative Events | | | | | |
| *Attempted Assassination Homeland* | 0.039 <br> (1.14) | 0.026 <br> (0.80) | | | | | 0.001 <br> (0.11) | 0.005 <br> (0.60) | | | | |
| *Attempted Assassination Imperial* | 0.059*** <br> (3.39) | 0.082*** <br> (7.25) | | | | | 0.013*** <br> (2.68) | 0.013*** <br> (2.92) | | | | |
| *Successful Assassination Homeland* | | | -0.028** <br> (2.07) | -0.003 <br> (0.19) | | | | | -0.004 <br> (1.18) | -0.002 <br> (1.42) | | |
| *Successful Assassination Imperial* | | | 0.027*** <br> (8.65) | -0.022 <br> (0.82) | | | | | 0.043*** <br> (2.90) | 0.025** <br> (2.31) | | |
| *Collective Unrest Homeland* | | | | | -0.007 <br> (0.26) | -0.013 <br> (1.24) | | | | | -0.003*** <br> (3.06) | -0.0003 <br> (2.06) |
| *Collective Unrest Imperial* | | | | | 0.008*** <br> (2.77) | 0.029*** <br> (2.64) | | | | | 0.012*** <br> (9.22) | 0.003*** <br> (2.74) |
| *Tsar Transition* | 0.064* <br> (1.86) | 0.099** <br> (1.96) | 0.088* <br> (1.89) | 0.085 <br> (1.53) | 0.053** <br> (2.39) | 0.121 <br> (1.64) | 0.086* <br> (1.76) | 0.105* <br> (1.85) | 0.141 <br> (1.51) | 0.119 <br> (1.40) | 0.036 <br> (1.22) | 0.083** <br> (2.09) |
| *Gold* | -0.009*** <br> (86.72) | -0.009*** <br> (3.03) | -0.008*** <br> (2.70) | -0.007* <br> (1.71) | -0.008*** <br> (3.33) | -0.014** <br> (2.38) | -0.008** <br> (2.50) | -0.009** <br> (2.19) | -0.010** <br> (2.17) | -0.009** <br> (2.31) | -0.005 <br> (0.90) | -0.007* <br> (1.75) |
| *Ruble-Guilder* | -0.819*** <br> (8.08) | -0.764*** <br> (5.47) | -0.527*** <br> (4.64) | -0.581*** <br> (3.62) | -0.534*** <br> (5.44) | 0.751*** <br> (5.26) | -0.536*** <br> (4.47) | -0.638*** <br> (4.41) | -1.346*** <br> (6.30) | -0.969*** <br> (5.82) | -1.121*** <br> (9.17) | -0.790*** <br> (4.92) |
| *Drought* | 0.005 <br> (1.20) | -0.005 <br> (0.90) | -0.006 <br> (0.91) | -0.006 <br> (1.00) | -0.001 <br> (0.35) | -0.001 <br> (0.20) | -0.005 <br> (0.84) | -0.004 <br> (0.57) | 0.001 <br> (0.07) | 0.001 <br> (0.18) | 0.022** <br> (2.29) | -0.003 <br> (0.52) |
| *Lagged DV* | 0.999*** <br> (124331.70) | 0.982*** <br> (219.39) | 0.989*** <br> (933.17) | 0.980*** <br> (235.12) | 0.991*** <br> (3771.63) | 1.012*** <br> (233.55) | 0.992*** <br> (6135.39) | 0.991*** <br> (361.21) | 0.996*** <br> (1954.13) | 1.018*** <br> (211.35) | 0.986*** <br> (2527.75) | 0.996*** <br> (155.43) |
| Constant | -0.008*** <br> (5.28) | 0.035*** <br> (4.58) | 0.046*** <br> (8.65) | 0.025*** <br> (3.96) | 0.040*** <br> (15.05) | -0.252*** <br> (4.39) | 0.033*** <br> (17.24) | 0.014*** <br> (3.08) | -0.351*** <br> (5.28) | -0.047*** <br> (4.74) | -0.378*** <br> (13.42) | -0.073* <br> (1.91) |
| Unrest Event Long-Term Volatility Homeland | 0.013 <br> (1.30) | 0.020* <br> (1.89) | 0.003 <br> (0.55) | -0.003 <br> (1.15) | 0.035 <br> (0.11) | 0.012 <br> (0.64) | 0.0005 <br> (0.80) | -0.008 <br> (1.34) | 0.001 <br> (0.57) | 0.003*** <br> (39.66) | -0.0001** <br> (1.99) | -0.001 <br> (1.58) |
| Unrest Event Short-Term Volatility Homeland | 0.007 <br> (0.39) | -0.017 <br> (1.17) | -0.005 <br> (0.84) | 0.004 <br> (0.88) | -0.035 <br> (0.11) | -0.02 <br> (0.68) | 0.0008 <br> (0.67) | 0.010 <br> (1.43) | -0.002 <br> (1.29) | -0.004** <br> (211.27) | 0.0004 <br> (1.60) | 0.002** <br> (1.97) |
| Unrest Event Long-Term Volatility Imperial | 0.003 <br> (0.98) | -0.003** <br> (2.06) | 0.020*** <br> (8.45) | 0.009 <br> (1.12) | 0.0003 <br> (0.01) | 0.001 <br> (0.05) | 0.0004 <br> (1.10) | 0.006 <br> (1.12) | -0.007 <br> (1.26) | -0.017* <br> (1.72) | 0.001*** <br> (5.65) | 0.0004 <br> (1.12) |
| Unrest Event Short-Term Volatility Imperial | -0.005 <br> (0.76) | -0.001 <br> (0.28) | -0.023*** <br> (6.54) | -0.012 <br> (1.19) | 0.0001 <br> (0.00) | 0.002 <br> (0.09) | -0.0005 <br> (0.65) | -0.006 <br> (1.09) | 0.01** <br> (2.39) | 0.020** <br> (1.97) | -0.0001 <br> (1.43) | -0.002 <br> (0.31) |
| GARCH-in-Mean | -0.016*** <br> (5.27) | -1.430*** <br> (2.69) | N/A | N/A | -0.088*** <br> (2.75) | -0.045** <br> (4.08) | -0.504** <br> (2.57) | -0.846* <br> (1.93) | -0.087*** <br> (5.28) | -0.880** <br> (6.11) | -0.091*** <br> (14.47) | -0.015** <br> (2.40) |
| Adjusted $R^2$ | 0.962 | 0.954 | 0.962 | 0.951 | 0.962 | 0.951 | 0.964 | 0.953 | 0.963 | 0.955 | 0.963 | 0.952 |
| AIC | -2.028 | -1.871 | -2.051 | -1.904 | -2.077 | -1.846 | -2.084 | -1.900 | -1.967 | -1.900 | -2.00 | -1.905 |
| n | 1125 | 1125 | 1125 | 1125 | 1125 | 1125 | 1125 | 1125 | 1125 | 1125 | 1125 | 1125 |

Table 8 reports results from asymmetric component GARCH-in-mean (ACGARCH-M) analyses of single unrest events and related cumulative unrest effects on sovereign risk in Russian homeland and imperial territories from 1820-1914. There are six focal unrest event types measured as 0-1 dummies taking the value of one when occurring in month *t*: individual unrest events in the forms of attempted assassinations in Russian homeland and imperial territories as well as successful assassinations in Russian homeland and imperial territories and collective unrest events of different forms (e.g., worker strikes) in Russian homeland and imperial territories. When these unrest event dummies take the value of one, we count the number of months in the previous year when the same type of unrest occurred anywhere in Russian homeland or imperial territories. We regress these 0-1 dummies in Columns 1-6 and these cumulative measures in Columns 7-12 on two measures of sovereign risk on these cumulative unrest indicators: month *t* 5-year Russian government bond yield percentages (yields) and month *t* 5-year Russian government bond spread percentage points above the benchmark 10-year British government bond yield percentage (spreads). Table 1 provides details on all variables included in these ACGARCH-M analyses. We report coefficient and volatility estimates along with absolute t-statistics (in parentheses) and levels of statistical significance for all right-hand side variables analyzed. GARCH-in-mean terms are insignificant in Columns 8-9 and not included in the model. Statistical significance levels include: *** = $p < 0.01$ (1% level), ** = $p < 0.05$ (5% level), * = $p < 0.10$ (10% level).

## Table 9: Results from ACGARCH-M analysis of single unrest events and sovereign risk in Russian homeland and imperial territories including spillovers (approach 1), 1820-1914

| **Dependent Variables →**<br>↓*Independent Variables* | **1**<br>**Yields** | **2**<br>**Spreads** | **3**<br>**Yields** | **4**<br>**Spreads** | **5**<br>**Yields** | **6**<br>**Spreads** |
|---|---|---|---|---|---|---|
| *Attempted Assassination Homeland* | 0.02<br>(0.51) | 0.015<br>(0.34) | | | | |
| *Attempted Assassination Imperial* | 0.062***<br>(3.61) | 0.057***<br>(2.94) | | | | |
| *Successful Assassination Homeland* | | | -0.026**<br>(2.28) | -0.002<br>(0.17) | | |
| *Successful Assassination Imperial* | | | 0.019<br>(0.88) | -0.019<br>(0.80) | | |
| *Collective Unrest Homeland* | | | | | -0.072***<br>(4.28) | -0.042***<br>(3.15) |
| *Collective Unrest Imperial* | | | | | 0.028***<br>(2.60) | 0.023*<br>(1.91) |
| *Multiple Events Dummy* | 0.018***<br>(2.90) | 0.019**<br>(2.63) | 0.019***<br>(3.11) | 0.025***<br>(2.88) | 0.035***<br>(3.14) | 0.048***<br>(3.15) |
| *Tsar Transition* | 0.082*<br>(1.98) | 0.081<br>(1.38) | 0.087*<br>(1.80) | 0.085<br>(1.28) | 0.113<br>(0.51) | 0.071<br>(1.12) |
| *Gold* | -0.009***<br>(3.36) | -0.009*<br>(1.97) | -0.008***<br>(2.66) | -0.007*<br>(1.68) | -0.008***<br>(3.29) | -0.015**<br>(2.41) |
| *Ruble-Guilder* | -0.537***<br>(4.36) | -0.569***<br>(3.73) | -0.512***<br>(3.93) | -0.676***<br>(4.63) | -1.121***<br>(5.88) | -0.884***<br>(4.66) |
| *Drought* | 0.0004<br>(0.06) | -0.004<br>(0.58) | -0.004<br>(0.64) | -0.002<br>(0.31) | -0.007<br>(0.69) | -0.011<br>(1.42) |
| *Lagged DV* | 0.992***<br>(1371.38) | 0.985***<br>(350.38) | 0.988***<br>(215.13) | 0.996***<br>(313.25) | 0.986***<br>(2485.32) | 0.983***<br>(4385.42) |
| Constant | -0.016<br>(0.87) | 0.014***<br>(3.18) | 0.047***<br>(193.89) | 0.011**<br>(2.37) | -0.343***<br>(14.98) | -0.239***<br>(4.09) |
| Unrest Event Long-Term Volatility Homeland | -0.007<br>(0.36) | 0.048<br>(0.19) | -0.012<br>(0.43) | -0.013<br>(0.56) | -0.003**<br>(2.31) | 0.015***<br>(2.59) |
| Unrest Event Short-Term Volatility Homeland | 0.019<br>(1.39) | -0.04<br>(0.16) | 0.01<br>(0.35) | 0.01<br>(0.52) | -0.005**<br>(2.51) | -0.019***<br>(2.67) |
| Unrest Event Long-Term Volatility Imperial | -0.001<br>(0.15) | -0.036<br>(0.20) | -0.02<br>(0.39) | -0.002<br>(0.16) | 0.002**<br>(2.34) | 0.001<br>(0.04) |
| Unrest Event Short-Term Volatility Imperial | 0.001<br>(0.30) | 0.036<br>(0.21) | 0.021<br>(0.39) | -0.0005<br>(0.03) | 0.001<br>(0.33) | 0.003<br>(0.011) |
| GARCH-in-Mean | -0.009***<br>(2.63) | N/A | N/A | -1.525***<br>(2.94) | -0.096***<br>(15.59) | -0.056***<br>(4.11) |
| Adjusted $R^2$ | 0.962 | 0.951 | 0.962 | 0.954 | 0.964 | 0.952 |
| AIC | -2.066 | -1.908 | -2.067 | -1.901 | -1.93 | -1.82 |
| *n* | 1124 | 1124 | 1124 | 1124 | 1124 | 1124 |

Table 9 reports results from asymmetric component GARCH-in-mean (ACGARCH-M) analyses of single unrest events effects on sovereign risk in Russian homeland and imperial territories from 1820-1914. To account for spillovers or contagion, a “multiple events” dummy is included to capture multiple categories of events happening in one month. Table 1 provides details on all variables included in these ACGARCH-M analyses. We report coefficient and volatility estimates along with absolute t-statistics (in parentheses) and levels of statistical significance for all right-hand side variables analyzed. Statistical significance levels include: *** = p < 0.01 (1% level), ** = p < 0.05 (5% level), * = p < 0.10 (10% level).

## Table 10: Results from ACGARCH-M analysis of single unrest events and sovereign risk in Russian homeland and imperial territories including spillovers (approach 2), 1820-1914

| Dependent Variables → <br> ↓*Independent Variables* | 1 <br> Yields | 2 <br> Spreads | 3 <br> Yields | 4 <br> Spreads | 5 <br> Yields | 6 <br> Spreads |
|---|---|---|---|---|---|---|
| *Attempted Assassination Homeland* | -0.039 <br> (1.42) | -0.004 <br> (0.14) | | | | |
| *Attempted Assassination Imperial* | 0.083** <br> (2.22) | 0.067** <br> (2.38) | | | | |
| *Successful Assassination Homeland* | | | -0.008 <br> (0.45) | -0.050*** <br> (2.77) | | |
| *Successful Assassination Imperial* | | | -0.66*** <br> (3.11) | -0.093** <br> (2.29) | | |
| *Collective Unrest Homeland* | | | | | -0.030*** <br> (2.82) | -0.034*** <br> (3.14) |
| *Collective Unrest Imperial* | | | | | 0.023** <br> (2.04) | 0.025*** <br> (2.67) |
| *Multiple Events Dummy Homeland* | -0.114*** <br> (3.51) | -0.090** <br> (2.05) | 0.051* <br> (1.71) | 0.042 <br> (1.37) | 0.012 <br> (0.56) | 0.028 <br> (1.33) |
| *Multiple Event Dummy Imperial* | 0.051*** <br> (2.88) | 0.032 <br> (0.96) | 0.092*** <br> (3.70) | 0.106*** <br> (3.94) | 0.034* <br> (1.65) | 0.025 <br> (1.52) |
| *Tsar Transition* | 0.109** <br> (2.47) | 0.021 <br> (0.63) | 0.032 <br> (0.60) | 0.054 <br> (1.27) | 0.071 <br> (1.0)7 | 0.095 <br> (0.96) |
| *Gold* | -0.010*** <br> (3.24) | -0.010** <br> (2.63) | -0.009*** <br> (2.69) | -0.010** <br> (2.14) | -0.015*** <br> (4.78) | -0.009* <br> (1.71) |
| *Ruble-Guilder* | -1.022*** <br> (10.62) | -0.926*** <br> (5.20) | -0.871*** <br> (6.89) | -1.125*** <br> (5.31) | -0.777*** <br> (4.60) | -1.198*** <br> (6.69) |
| *Drought* | .008* <br> (1.83) | 0.001 <br> (0.08) | 0.003 <br> (0.47) | 0.005 <br> (0.61) | 0.0004 <br> (0.42) | 0.002 <br> (0.29) |
| *Lagged DV* | 0.996*** <br> (2271.81) | 1.012*** <br> (103.98) | 1.001*** <br> (685.99) | 1.023*** <br> (112.83) | 0.988*** <br> (1406.01) | 1.001*** <br> (173.53) |
| Constant | -0.132*** <br> (52.90) | -0.147** <br> (2.45) | -0.10*** <br> (3.86) | -0.283*** <br> (3.81) | 0.143*** <br> (9.76) | -0.210*** <br> (3.61) |
| Unrest Event Long-Term Volatility Homeland | 0.008 <br> (1.33) | 0.116 <br> (0.72) | -0.001 <br> (0.22) | -0.018*** <br> (3.52) | 0.021*** <br> (10.03) | 0.059 <br> (1.35) |
| Unrest Event Short-Term Volatility Homeland | 0.014 <br> (0.78) | -0.119 <br> (0.74) | 0.002 <br> (0.66) | 0.015*** <br> (3.45) | -0.025*** <br> (18.35) | -0.097 <br> (1.38) |
| Unrest Event Long-Term Volatility Imperial | -0.003** <br> (2.10) | 0.016 <br> (0.17) | -0.008 <br> (1.03) | -0.006** <br> (2.44) | 0.001 <br> (0.04) | -0.003 <br> (0.03) |
| Unrest Event Short-Term Volatility Imperial | 0.004 <br> (1.20) | -0.014 <br> (0.14) | -0.002*** <br> (5.14) | 0.003 <br> (0.99) | 0.002 <br> (0.07) | 0.006 <br> (0.06) |
| GARCH-in-Mean | -0.028*** <br> (57.93) | -0.026*** <br> (2.63) | -0.017*** <br> (4.08) | -0.052*** <br> (3.77) | -0.0995*** <br> (8.74) | -0.041*** <br> (3.90) |
| Adjusted $R^2$ | 0.962 | 0.951 | 0.963 | 0.952 | 0.964 | 0.952 |
| AIC | -2.031 | -1.898 | -2.051 | -1.893 | -1.99 | -1.859 |
| n | 1124 | 1124 | 1124 | 1124 | 1124 | 1124 |

Table 10 reports results from asymmetric component GARCH-in-mean (ACGARCH-M) analyses of single unrest events effects on sovereign risk in Russian homeland and imperial territories from 1820-1914. To account for spillovers or contagion, two interaction dummies are included, equal to 1 if the category in question is joined by an event of a different category in the same month. Table 1 provides details on all variables included in these ACGARCH-M analyses. We report coefficient and volatility estimates along with absolute t-statistics (in parentheses) and levels of statistical significance for all right-hand side variables analyzed. Statistical significance levels include: *** = $p < 0.01$ (1% level), ** = $p < 0.05$ (5% level), * = $p < 0.10$ (10% level).

**Table 11: Results from Heckman analysis of single event unrest and sovereign risk in Russian homeland and imperial territories, 1820-1914**

| **Dependent Variables →** | **Yields** | | | **Spreads** | | |
|---|---|---|---|---|---|---|
| | **1** | **2** | **3** | **4** | **5** | **6** |
| ↓Independent Variables | **Heckman ML** | **Heckman ML** | **Heckman ML** | **Heckman ML** | **Heckman ML** | **Heckman ML** |
| *Distance* (from St. Petersburg to Unrest) | 0.0002***<br>(3.61) | 0.0002***<br>(3.46) | 0.0003***<br>(4.85) | 0.0001**<br>(2.29) | 0.0001**<br>(2.10) | 0.0003***<br>(4.63) |
| *Bond Volatility* | | 18.532***<br>(5.06) | 17.307***<br>(6.54) | | 16.022***<br>(3.95) | 14.799***<br>(4.82) |
| *Years Under Russian Rule* | | | -0.002<br>(0.75) | | | 0.0002<br>(0.72) |
| *Distance*Years Under Russian Rule* | | | -0.000007***<br>(2.77) | | | -0.00001***<br>(3.53) |
| *Tsar Transition* | 0.222**<br>(2.03) | 0.242**<br>(2.25) | 0.258**<br>(2.07) | 0.367<br>(0.17) | 0.054<br>(0.25) | 0.079<br>(0.32) |
| *Gold* | 0.128<br>(0.73) | 0.125<br>(0.76) | 0.016<br>(1.08) | 0.009<br>(0.62) | 0.008<br>(0.65) | 0.009<br>(0.71) |
| *Ruble-Guilder* | -4.021***<br>(3.05) | -3.396***<br>(2.73) | -3.633***<br>(3.12) | -4.117***<br>(2.91) | -3.580***<br>(2.68) | -3.134***<br>(2.77) |
| *Serfdom* | -0.450***<br>(4.91) | -0.453***<br>(5.13) | -0.393***<br>(4.45) | -0.582***<br>(8.37) | -0.585***<br>(8.73) | -0.500***<br>(7.76) |
| *Drought Intensity* | -0.017<br>(0.16) | -0.159**<br>(2.31) | -0.121*<br>(1.95) | -0.110<br>(0.12) | -0.134**<br>(2.42) | -0.121**<br>(2.42) |
| *Cereals* | 0.001<br>(0.00) | -0.002<br>(0.03) | -0.029<br>(0.50) | -0.003<br>(0.06) | -0.005<br>(0.10) | -0.001<br>(0.03) |
| *Interior* | -17.44*<br>(1.88) | -21.339**<br>(2.36) | -33.951***<br>(2.71) | 6.255<br>(1.05) | 2.883<br>(0.51) | -19.403**<br>(2.28) |
| Constant | 5.017***<br>(6.45) | 5.067***<br>(6.84) | 5.327***<br>(7.99) | 1.933***<br>(3.31) | 1.978***<br>(3.60) | 1.824***<br>(3.35) |
| Inverse Mills Ratio | -0.262**<br>(2.33) | -0.286***<br>(2.80) | -0.158<br>(1.33) | -0.184<br>(1.56) | -0.205**<br>(2.00) | -0.230**<br>(2.01) |
| Selected n | 232 | 232 | 199 | 232 | 232 | 199 |
| *n* | 1019 | 1019 | 986 | 1019 | 1019 | 986 |

Table 11 reports results from second-stage Heckman regressions using Heckman selection methods on two measures of sovereign risk on these cumulative unrest indicators: month *t* 5-year Russian government bond yield percentages (yields) and month *t* 5-year Russian government bond spread percentage points above the benchmark 10-year British government bond yield percentage (spreads). Right-hand side variables in the second stage other than distance are included in the first-stage selection equation. Table 1 provides details on all variables included in these analyses. We report coefficient and volatility estimates along with absolute t-statistics (in parentheses) and levels of statistical significance for all right-hand side variables analyzed. Statistical significance levels include: *** = $p < 0.01$ (1% level), ** = $p < 0.05$ (5% level), * = $p < 0.10$ (10% level).

**Table 12: Results from ACGARCH-M analysis of single unrest events and sovereign risk in Russian homeland territories, imperial territories of modern-day Ukraine, and other imperial territories, 1820-1914**

| **Dependent Variables →**<br><br>↓*Independent Variables* | **1**<br><br>**Yields** | **2**<br><br>**Spreads** | **3**<br><br>**Yields** | **4**<br><br>**Spreads** |
|---|---|---|---|---|
| *Collective Unrest Homeland* | -0.029**<br>(2.49) | -0.018*<br>(1.93) | -0.028***<br>(3.73( | -0.016<br>(1.60) |
| *Collective Unrest Ukraine* | 0.067***<br>(5.98) | 0.108***<br>(5.37) | 0.051***<br>(2.62) | 0.069***<br>(3.06) |
| *Collective Unrest Other Imperial Territories* | | | 0.020*<br>(1.87) | 0.013<br>(1.24) |
| *Tsar Transition* | 0.020<br>(0.28) | 0.056<br>(1.46) | 0.118<br>(1.28) | 0.092<br>(1.08) |
| *Gold* | -0.011**<br>(2.10) | -0.012*<br>(1.81) | -0.009*<br>(1.84) | -0.009**<br>(2.28) |
| *Ruble-Guilder* | -1.142***<br>(8.85) | -1.338***<br>(13.64) | -0.805***<br>(5.41) | -1.054***<br>(5.26) |
| *Drought* | 0.006<br>(0.57) | -0.004<br>(1.13) | 0.007<br>(0.76) | 0.005<br>(0.54) |
| *Lagged DV* | 0.991***<br>(1208.26) | 1.046***<br>(5313.78) | 0.987***<br>(1068.12) | 1.020***<br>(106.76) |
| Constant | -0.377***<br>(6.93) | 0.068***<br>(15.02) | 0.145***<br>(9.55) | -0.251***<br>(3.30) |
| ↓*GARCH Attributes* | | | | |
| Unrest Event Long-Term Volatility Homeland | 0.020***<br>(2.83) | -0.004*<br>(1.75) | 0.018***<br>(3.16) | -0.005<br>(0.99) |
| Unrest Event Short-Term Volatility Homeland | -0.025***<br>(3.20) | 0.002<br>(0.81) | -0.023***<br>(3.80) | 0.003<br>(0.56) |
| Unrest Event Long-Term Volatility Ukraine | 0.005<br>(0.07) | -0.026**<br>(2.21) | 0.002<br>(0.03) | 0.005<br>(0.16) |
| Unrest Event Short-Term Volatility Ukraine | 0.004<br>(0.07) | 0.037***<br>(2.86) | 0.005<br>(0.08) | 0.004<br>(0.14) |
| Unrest Event Long-Term Volatility, Other Imperial Territories | | | 0.0006<br>(0.02) | -0.001<br>(0.17) |
| Unrest Event Short-Term Volatility, Other Imperial Territories | | | 0.002 | 0.002 |
| | | | (0.08) | (0.24) |
| GARCH-in-Mean | -0.091***<br>(7.49) | -1.548***<br>(30.05) | -0.967***<br>(6.36) | -0.045***<br>(3.41) |
| Adjusted $R^2$ | 0.963 | 0.955 | 0.964 | 0.952 |
| AIC | -1.984 | -1.879 | -1.989 | -1.894 |
| *n* | 1125 | 1125 | 1124 | 1124 |

Columns 1-2 of Table 12 report results from asymmetric component GARCH-in-mean (ACGARCH-M) analyses of single collective unrest event effects on sovereign risk in Russian homeland and imperial territories comprising modern-day Ukraine from 1820-1914. Columns 3-4 report the same results for single collective unrest events in imperial territories of modern-day Ukraine versus other imperial territories from 1820-1914. In Columns 1-2, there are two unrest event types measured as 0-1 dummies taking the value of one when occurring in month *t*:: collective unrest events of different forms (e.g., worker strikes) in Russian homeland territories and imperial territories of modern-day Ukraine. In Columns 3-4, there again is the 0-1 dummy for collective unrest events in imperial territories of modern-day Ukraine and another 0-1 dummy taking the value of one when collective unrest events occur in other imperial territories. We regress two measures of sovereign risk on these single collective unrest indicators: month *t* 5-year Russian government bond yield percentages (yields) and month *t* 5-year Russian government bond spread percentage points above the benchmark 10-year British government bond yield percentage (spreads). Table 1 provides details on the control variables included in these ACGARCH-M analyses. We report coefficient and volatility estimates along with the absolute value of t-statistics (in parentheses) and levels of statistical significance for all right-hand side variables analyzed. Statistical significance levels include: *** = $p < 0.01$ (1% level), ** = $p < 0.05$ (5% level), * = $p < 0.10$ (10% level).

## Table 13: Event Study Results, Unrest in the Imperial Territory of Ukraine

| | | Raw Returns Model | | Constant Mean Model | |
|---|---|---|---|---|---|
| **Bond Yield Returns** | | [-1,+1] | | [-1,+1] | |
| | **n** | **CAAR** | | | |
| | | **Patell Adjusted** | **GRANKT** | **Patell Adjusted** | **GRANKT** |
| Unrest Imperial, Ukraine Only | 43 | 0.008 | | 0.011 | |
| | | *2.41*** | *0.97* | *3.41**** | *0.39* |
| | | [-4,+4] | | [-4,+4] | |
| | | **Raw Returns Model** | | **Constant Mean Model** | |
| | 43 | **Patell Adjusted** | **GRANKT** | **Patell Adjusted** | **GRANKT** |
| | | 0.028 | | 0.037 | |
| | | *7.99**** | *4.42**** | *10.03**** | *6.09**** |

Table 13 reports results from event study analyses of abnormal 5-year Russian government bond yield percentage returns (yield returns) for each category of political unrest. Event study output includes yield returns for one month prior and one month following the event in the first rows and four months prior and four months following in the second panel. Model run using both raw returns and constant mean returns as baseline for calculating abnormal returns. Patell Adjusted is the parametric test of Kolari and Pynnönen (2010) while GRANKT is the generalized rank t-test of Kolari and Pynnönen (2011). Absolute values of t-stats shown in italics at statistical significance levels: *** = $p < 0.01$ (1% level), ** = $p < 0.05$ (5% level), * = $p < 0.10$ (10% level).

**Internet Appendix**

**Additional Robustness Tests**

*Sub-Sampling and Unrest Effects in Early Years*

ACGARCH-M results in Table 4 consistently indicate that location matters for understanding the impact of unrest on sovereign risk. That understanding aligns with the projection rather than proximity perspective. However, these results raise a follow-on issue related to sampling, as it could be that ACGARCH-M models assuming contemporaneous unrest effects on sovereign risk are not properly specified, at least in earlier years of our study. The growth of railroad, telegraph, and foreign correspondent networks by the mid-19th century meant that information regarding unrest events was reaching foreign financial markets in a matter of hours or days rather than weeks. But those networks did not exist in December 1825 when troops threatened to topple a newly proclaimed Tsar Nicholas I in St. Petersburg. We treated these early unrest events as we did unrest events in the mid-19th or early 20th centuries, occurring in month *t* and affecting Russian government bond yields and spreads in the same month *t*. Perhaps after lagging their effects consistent with the slower flow of information to foreign financial markets, results for such early unrest events would disappear.

To investigate this possibility, we re-estimate ACGARCH-M models with a sub-sample of 105 collective unrest events occurring from 1820-1855, with 64 occurring in Russian homeland territories and 41 in imperial territories. We lag the *Unrest* dummy by a month ($t - 1$). This widens the window of information transmission to foreign financial markets to about 60 days, more time than a "worst-case" scenario for correspondence by sailing ship from St. Petersburg to London in the late 18th century (Vinnal, 2014).

Results in Table a.1 are consistent with earlier trends but with one difference. Columns 3-4 show confirmation of the results for Russian government bond yields, where coefficients on single collective unrest events (0.080, $p < 0.01$) and cumulative collective unrest (0.010, $p < 0.05$) in imperial territories are positive and significantly different from zero and also different from the coefficients on Russian homeland unrest. However, when looking at Column 2, spreads on Russian government bonds are positive and significant at the one percent level (0.004, $p < 0.001$) for cumulative collective unrest in imperial territories but the coefficient for cumulative unrest in Russian homeland territories is also significant and positive (0.012). Indeed, this is the only time when unrest in the Russian homeland is both positive and significant and greater than the effects from the imperial territories and should be taken with a grain of salt: this model had many issues converging and the AIC (-1.35) is much higher than other models throughout the paper. Thus, it may not be representative of the "true" state of affairs, or it may just be a quirk of the 1820 to 1855 timeframe. But overall, when lagged for potentially slower communications and transportation, early-period unrest events increase sovereign risk – as proxied by bond yields – more in imperial territories again consistent with the projection perspective.

As an additional test, Table A.2 shows the sample for post-1855 (i.e., from January 1856 onward) without the month lag, and the results corroborate both the whole sample results from Table 4 and the results for pre-1855 shown in Table A.1. Unrest in the imperial territories is far more likely to raise both Russian government bond yields and risk spreads than unrest occurring in the Russian homeland, with yields in particular moving as much as 5 basis points for a single unrest event (Column 3). Interestingly, there appear to be no effects on the volatility of either spreads or yields, but in each model, the GARCH-M term is highly economically and statistically significant, meaning that any volatility which is caused by political unrest has its greatest impact on the conditional mean.

*Unrest in Muscovy and Onward*

Similar to the robustness test shown in the main text on Ukraine and other imperial (non-Ukrainian) territories, we also explore here whether unrest in Russia's historical Muscovy Duchy affected yields and spreads differently. We use here the Muscovy Duchy as it was in 1505, including the Moscow region and the area surrounding Perm (added in 1478) but excluding St. Petersburg and all of the southern Volga region. As with the Ukraine robustness test, we have thus split up the unrest dummies into unrest in the historic Muscovy Duchy, unrest in the rest of the Russian homeland as of 1820, and unrest occurring in the imperial territories. The results are shown in Table A.3 for both yields and spreads and contain some interesting effects: for bond yields, unrest in the imperial territories continues to show significant (at the two percent level) effects, with an increase in yields of approximately 1.5 basis points per unrest event. Unrest in the Muscovy Duchy is insignificant while unrest elsewhere in the Russian homeland results in a statistically significant (at the five percent level) 1.4 average basis point drop in yields. We conjecture that the differential effect here is one of power projection in already-consolidated territories, where calling in the army is judged by markets to more likely be successful than in far-flung and remote territories still being subjugated. This can be further confirmed by the statistically significant (at the 10 percent level) drop in volatility over the long-term surrounding events which occur throughout the Russian homeland.

A similar effect holds when looking at spreads, although the effect of unrest in the Duchy and in the Russian homeland disappears in significance. Most importantly, unrest in imperial territories increases the spread of Russian bonds over British bonds on average by 0.6 basis points, a result significant at the two percent level. As one of the few models where volatility is not important in the conditional mean, it appears that the effect of unrest manifests itself in the conditional variance: in particular, unrest in the imperial territories has a significant ($p=0.0265$) increase on spread volatility in the long run, while unrest in Russia outside of Muscovy has a severe dampening effect on volatility in the short run (coefficient of -0.002, $p<0.01$). This is likely due to the same effects as observed in the conditional mean in the bond yield equation, related to our conjecture above.

*Endogeneity of Unrest*

In addition to the sample selection issues in the Heckman equation shown in (8), distance could itself be endogenously determined, in that the location of unrest can be dependent on currently unobserved factors. That is, it is entirely plausible that the location of unrest itself is determined by several endogenous factors, which then translates into increased risk for the sovereign. To account for this possibility, we have taken Equation 8 and re-run it as an instrumental variable-generalized method of moments (IV-GMM) approach. More specifically, we instrument for *Distance* with three terms plausibly correlated with unrest distance from St. Petersburg and taken directly from the political science literature, but plausibly uncorrelated with sovereign risk: local (oblast or city) size in square kilometers (*Size*), which should be negatively related to distance from St. Petersburg; local population density (*Density*) as the number of individuals per square kilometer, which should be negatively related to distance from St. Petersburg; and a 0-1 dummy taking the value of one if Russia had lost a war in the previous month (*Lost War*). The *Lost War* dummy, by definition, occurs far from St. Petersburg and could plausibly encourage unrest in and around contested peripheral regions.

Results from the IV-GMM estimation appear in Table A.4 (run both without and with bond volatility included), and they again support the projection perspective. Indeed, when instrumenting for the potential endogeneity of the unrest location, coefficient magnitudes for *Distance* increase five times for 5-year Russian government bond yields (0.0005, $p < 0.01$) compared to Heckman ML estimation without instrumentation (0.0001, $p < 0.001$) and four times for spreads (0.0004, $p < 0.01$) compared to Heckman ML estimation without instrumentation (0.0001, $p < 0.01$). When bond volatility is included (Columns 2 and 4), similar magnitudes are obtained, with distance increasing yields by four times over the ML estimation and risk spreads by three times. These results are supported by commonly used GMM diagnostics, such as the Kleibergen-Papp test, which rejects the null of model under-identification and Hansen's *J* test, which does not reject the null hypothesis that the instruments are valid as a group. Using the volatility model as a baseline, we can say that each thousand kilometers of distance from St. Petersburg, more distant unrest raises bond yields by approximately 40 basis points and spreads by 30 basis points. The price of empire increases as imperial boundaries expand and unrest associated with those expanding boundaries emerges.

## APPENDIX REFERENCE

**Table A.1: Results from ACGARCH-M analysis of unrest and sovereign risk in Russian homeland and imperial territories, 1820-1855**

| **Dependent Variables →** | **Spreads** | | **Yields** | |
|---|---|---|---|---|
| ↓*Independent Variables* | 1 | 2 | 3 | 4 |
| | | | | |
| *Lagged Collective Unrest Homeland* | -0.009<br>(0.79) | | -0.0005<br>(0.02) | |
| *Lagged Collective Unrest Imperial* | -0.009<br>(0.85) | | 0.080***<br>(2.62) | |
| *Lagged Cumulative Collective Unrest Homeland* | | 0.012***<br>(9.42) | | -0.004<br>(1.04) |
| *Lagged Cumulative Collective Unrest Imperial* | | 0.004***<br>(4.25) | | 0.010**<br>(2.30) |
| ↓*GARCH Attributes* | | | | |
| Unrest Event Long-Term Volatility Homeland | 0.001<br>(0.79) | 0.030<br>(1.45) | -0.008<br>(0.71) | 0.008<br>(0.50) |
| Unrest Event Long-Term Volatility Imperial | 0.009*<br>(1.86) | -0.003<br>(0.09) | -0.001<br>(0.05) | 0.0005<br>(0.05) |
| Unrest Event Short-Term Volatility Homeland | -0.002<br>(0.84) | -0.007<br>(0.32) | 0.009<br>(0.73) | -0.006<br>(0.41) |
| Unrest Event Short-Term Volatility Imperial | -0.010*<br>(1.87) | -0.005<br>(0.15) | 0.01<br>(0.33) | 0.002<br>(0.22) |
| GARCH-in-Mean | 0.030***<br>(4.00) | -0.038***<br>(12.53) | -0.120***<br>(3.83) | -0.140**<br>(2.52) |
| Adjusted $R^2$ | 0.699 | 0.699 | 0.869 | 0.678 |
| AIC | -1.96 | -1.35 | -2.17 | -2.15 |
| *N* (Event *N*) | 422 (105) | 422 (105) | 422 (105) | 422 (105) |

Table A.1 reports results from asymmetric component GARCH-in-mean (ACGARCH-M) analyses of single unrest event and related cumulative unrest effects on sovereign risk in Russian homeland and imperial territories from 1820-1855. We regress two measures of sovereign risk on these unrest indicators: month *t* 5-year Russian government bond yield percentages (yields) and month *t* 5-year Russian government bond spread percentage points above the benchmark 10-year British government bond yield percentage (spreads). Table 1 provides details on all variables included in these ACGARCH-M analyses. Controls not reported but available from the authors. We report coefficient and volatility estimates along with the absolute value of t-statistics below coefficients. Statistical significance levels include: *** = p < 0.01 (1% level), ** = p < 0.05 (5% level), * = p < 0.10 (10% level).

**Table A.2: Results from ACGARCH-M analysis of unrest and sovereign risk in Russian homeland and imperial territories, 1856-1914**

| **Dependent Variables →** | **Spreads** | | **Yields** | |
|---|---|---|---|---|
| ↓*Independent Variables* | 1 | 2 | 3 | 4 |
| | | | | |
| *Collective Unrest Homeland* | 0.006<br>(0.28) | | -0.008<br>(0.49) | |
| *Collective Unrest Imperial* | 0.039***<br>(2.86) | | 0.050***<br>(4.70) | |
| *Cumulative Collective Unrest Homeland* | | -0.004<br>(1.57) | | -0.015***<br>(3.00) |
| *Cumulative Collective Unrest Imperial* | | 0.007***<br>(3.01) | | 0.014***<br>(3.18) |
| ↓*GARCH Attributes* | | | | |
| Unrest Event Long-Term Volatility Homeland | 0.061<br>(0.67) | 0.022<br>(0.26) | 0.02<br>(1.17) | 0.004<br>(0.55) |
| Unrest Event Long-Term Volatility Imperial | 0.017<br>(0.35) | -0.004<br>(0.04) | 0.001<br>(0.11) | -0.002<br>(0.04) |
| Unrest Event Short-Term Volatility Homeland | -0.056<br>(0.63) | -0.020<br>(0.24) | -0.030<br>(1.43) | -0.004<br>(0.68) |
| Unrest Event Short-Term Volatility Imperial | -0.0151<br>(0.33) | 0.001<br>(0.09) | 0.002<br>(0.40) | 0.001<br>(0.34) |
| GARCH-in-Mean | -3.745**<br>(2.43) | -0.040***<br>(19.80) | -4.491***<br>(3.66) | -1.360***<br>(2.90) |
| Adjusted $R^2$ | 0.972 | 0.971 | 0.979 | 0.979 |
| AIC | -1.839 | -1.835 | -1.848 | -1.858 |
| *N* | 702 | 702 | 702 | 702 |

Table A.2 reports results from asymmetric component GARCH-in-mean (ACGARCH-M) analyses of single unrest event and related cumulative unrest effects on sovereign risk in Russian homeland and imperial territories from 1856 to 1914. We regress two measures of sovereign risk on these unrest indicators: month *t* 5-year Russian government bond yield percentages (yields) and month *t* 5-year Russian government bond spread percentage points above the benchmark 10-year British government bond yield percentage (spreads). Table 1 provides details on all variables included in these ACGARCH-M analyses. Controls not reported but available from the authors. We report coefficient and volatility estimates along with the absolute value of t-statistics below coefficients. Statistical significance levels include: *** = p < 0.01 (1% level), ** = p < 0.05 (5% level), * = p < 0.10 (10% level).

**Table A.3: Results from ACGARCH-M analysis of unrest and sovereign risk in the historic Muscovy Duchy, the Russian homeland, and imperial territories, 1820-1914**

| **Dependent Variables →**<br>↓*Independent Variables* | **1**<br>**Yields** | **2**<br>**Spreads** |
|---|---|---|
| *Collective Unrest, Muscovy Duchy* | -0.03<br>(0.37) | 0.006<br>(0.77) |
| *Collective Unrest, Elsewhere in Russian Homeland* | -0.014**<br>(2.08) | -0.0001<br>(0.02) |
| *Collective Unrest, Imperial Territories* | 0.015**<br>(2.53) | 0.006**<br>(2.41) |
| *Tsar Transition* | 0.080<br>(1.43) | 0.084<br>(1.33) |
| *Gold* | -0.009***<br>(2.77) | -0.007*<br>(1.69) |
| *Ruble-Guilder* | -0.773***<br>(5.88) | -0.628***<br>(4.03) |
| *Drought* | -0.009*<br>(1.74) | -0.003<br>(0.39) |
| *Lagged Dependent Variable* | 1.00***<br>(524.79) | 0.98***<br>(251.48) |
| Constant | 0.033***<br>(6.59) | 0.017***<br>(2.85) |
| ↓*GARCH Attributes* | | |
| *Unrest Event Long-Term Volatility, Muscovy* | -0.003<br>(0.53) | -0.0002<br>(0.65) |
| *Unrest Event Short-Term Volatility, Muscovy* | 0.0001<br>(0.12) | -0.0007<br>(0.56) |
| *Unrest Event Long-Term Volatility, Homeland* | -0.001*<br>(1.82) | -0.0005*<br>(1.94) |
| *Unrest Event Short-Term Volatility, Homeland* | -0.001<br>(0.73) | -0.0002***<br>(2.71) |
| *Unrest Event Long-Term Volatility, Imperial* | 0.0005<br>(0.17) | 0.0005**<br>(2.22) |
| *Unrest Event Short-Term Volatility, Imperial* | 0.001<br>(0.55) | -0.001<br>(0.53) |
| GARCH-in-Mean | -0.267***<br>(3.76) | n/a |
| Adjusted $R^2$ | 0.963 | 0.951 |
| AIC | -2.056 | -1.904 |
| N | 1124 | 1124 |

Columns 1-2 of Table A.3 reports results from asymmetric component GARCH-in-mean (ACGARCH-M) analyses of single collective unrest event effects on sovereign risk in the historic Muscovy Duchy, the Russian homeland, and imperial territories from 1820-1914. Shown are three 0-1 dummies for collective unrest events in imperial territories, the Russian homeland outside of the Muscovy Duchy as of 1505, and unrest which occurred within the borders of the Duchy from 1505. We regress two measures of sovereign risk on these single collective unrest indicators: month *t* 5-year Russian government bond yield percentages (yields) and month *t* 5-year Russian government bond spread percentage points above the benchmark 10-year British government bond yield percentage (spreads). Table 1 provides details on the control variables included in these ACGARCH-M analyses. We report coefficient and volatility estimates along with the absolute value of the t-statistics (in parentheses) and the levels of statistical significance for all right-hand side variables analyzed. Statistical significance levels include: *** = $p < 0.01$ (1% level), ** = $p < 0.05$ (5% level), * = $p < 0.10$ (10% level).

## Table A.4 – Heckman Models with Unrest as an Endogenous Variable

| Dependent Variables → | Yields | | Spreads | |
|---|---|---|---|---|
| | 1 | 2 | 3 | 4 |
| ↓Independent Variables | IV-GMM | IV-GMM | IV-GMM | IV-GMM |
| *Distance* (from St. Petersburg to Unrest) | 0.0005***<br>-3.82 | 0.0001***<br>-3.62 | 0.0004***<br>-3.53 | 0.0003***<br>-3.33 |
| *Bond Volatility* | | 15.431***<br>-3.72 | | 13.140***<br>-3.6 |
| *Tsar Transition* | 0.551***<br>-3.92 | 0.531***<br>-4.06 | 0.275**<br>-2.36 | 0.259**<br>-2.08 |
| *Gold* | 0.003<br>-0.12 | 0.001<br>-0.05 | 0.005<br>-0.26 | 0.003<br>-0.18 |
| *Ruble-Guilder* | -1.695<br>-0.94 | -1.398<br>-0.96 | -1.574<br>-0.96 | -1.338<br>-0.99 |
| *Serfdom* | -0.419**<br>-2.05 | -0.439**<br>-2.25 | -0.561***<br>-4.38 | -0.580***<br>-4.69 |
| *Drought Intensity* | 0.039<br>-0.35 | -0.071<br>-0.76 | 0.036<br>-0.56 | -0.055<br>-1.01 |
| *Cereals* | -0.009<br>-0.08 | -0.008<br>-0.08 | -0.006<br>-0.09 | -0.006<br>-0.09 |
| *Interior* | 7.852<br>-0.44 | 2.279<br>-0.13 | 23.914**<br>-2.12 | 19.195*<br>-1.79 |
| Constant | 4.695***<br>-3.64 | 4.797***<br>-3.95 | 1.594*<br>-1.84 | 1.683**<br>-2.07 |
| Inverse Mills Ratio | -0.207**<br>-2.49 | -0.229***<br>-2.8 | -0.107<br>-1.58 | -0.128**<br>-1.97 |
| Under-Identification (Kleibergen-Papp) Test (p-value) | 0.005 | 0.005 | 0.001 | 0.002 |
| Instrument Exogeneity (Hansen J) Test (p-value) | 0.196 | 0.196 | 0.224 | 0.232 |
| Selected *N* | 311 | 311 | 311 | 311 |
| *N* | 311 | 311 | 311 | 311 |

Table A.4 reports results from second-stage Heckman IV-GMM regressions using Heckman selection methods on two measures of sovereign risk on these cumulative unrest indicators: month *t* 5-year Russian government bond yield percentages (yields) and month *t* 5-year Russian government bond spread percentage points above the benchmark 10-year British government bond yield percentage (spreads). Right-hand side variables in the second stage other than distance are included in the first-stage selection equation. The IV-GMM analysis shown in Columns 2 and 4 include three instruments for *Distance*: local size (*Size*); local population density (*Density*); and whether Russia lost a war in the previous month (*Lost War*). Table 1 provides details on all variables included in these analyses. We report coefficient and volatility estimates along with absolute t-statistics (in parentheses) and levels of statistical significance for all right-hand side variables analyzed. Statistical significance levels include: *** = p < 0.01 (1% level), ** = p < 0.05 (5% level), * = p < 0.10 (10% level).